\documentclass[a4paper, 11pt]{article}
\pdfoutput=1

\usepackage{a4wide}

\usepackage[utf8]{inputenc}
\usepackage[T1]{fontenc}
\usepackage{lmodern}
\usepackage{textcomp}
\usepackage{microtype}

\usepackage[english]{babel}
\usepackage[autostyle, english = american]{csquotes}
\MakeOuterQuote{"}

\usepackage{mathtools}
\usepackage{upgreek}
\usepackage{units}
\usepackage{slashed}
\usepackage{physics}

\allowdisplaybreaks
\DeclareMathOperator{\diag}{diag}
\DeclareMathOperator{\BR}{BR}
\DeclareMathOperator{\iu}{i}

\makeatletter
\g@addto@macro\bfseries{\boldmath}
\makeatother

\usepackage{pgfplots}
\DeclareUnicodeCharacter{FEFF}{~}
\usetikzlibrary{external}

\usepackage{footnote}
\usepackage{subcaption}
\usepackage{booktabs}
\usepackage{multirow}
\usepackage{graphicx}

\graphicspath{{./Figures/}}

\usepackage[sorting = none,style = numeric-comp, giveninits = true, backend = bibtex]{biblatex}
\bibliography{bibliography}

\usepackage[hidelinks]{hyperref}

\usepackage{fancyhdr}
\fancypagestyle{title}{
  \setlength{\headheight}{23pt}
  
  \fancyhead[R]{\small CP3-18-02\\TUM-HEP-1128/18}
}

\usepackage{authblk}

\title{NA62 sensitivity to heavy neutral leptons\\in the low scale seesaw model}

\author[1,2]{Marco Drewes}
\author[3,1]{Jan Hajer}
\author[4,2]{Juraj Klaric}
\author[5]{Gaia Lanfranchi}

\affil[1]{Centre for Cosmology, Particle Physics and Phenomenology,\protect\\Université catholique de Louvain, Louvain-la-Neuve B-1348, Belgium}
\affil[2]{Excellence Cluster Universe, Boltzmannstraße 2, D-85748, Garching, Germany}
\affil[3]{Institute for Advanced Study, The Hong Kong University of Science and Technology,\protect\\Clear Water Bay, Kowloon, Hong Kong S.A.R., China}
\affil[4]{Physik Department T70, Technische Universität München,\protect\\James Franck Straße 1, D-85748 Garching, Germany}
\affil[5]{Laboratori Nazionali INFN di Frascati, Frascati, Italy}

\date{}

\begin{document}

\maketitle
\thispagestyle{title}

\begin{abstract}
\noindent The sensitivity of beam dump experiments to heavy neutral leptons depends on the relative strength of their couplings to individual lepton flavours in the Standard Model.
We study the impact of present neutrino oscillation data on these couplings in the minimal type I seesaw model and find that it significantly constrains the allowed heavy neutrino flavour mixing patterns.
We estimate the effect that the DUNE experiment will have on these predictions.
We then discuss implication that this has for the sensitivity of the NA62 experiment when operated in the beam dump mode and provide sensitivity estimates for different benchmark scenarios.
We find that the sensitivity can vary by almost two orders of magnitude for general choices of the model parameters, but depends only weakly on the flavour mixing pattern within the parameter range that is preferred by neutrino oscillation data.
\end{abstract}

\begin{small}
\tableofcontents
\end{small}

\section{Introduction}

All fermions in the Standard Model (SM) of particle physics with the exception of neutrinos are known to exist with both, left handed and right handed chirality.
A particularly strong motivation for the existence of right handed neutrinos $\nu_R$ comes from the fact that they can explain the light neutrino flavour oscillations via the \emph{type I seesaw mechanism}~\cite{Minkowski:1977sc, GellMann:1980vs, Mohapatra:1979ia, Yanagida:1980xy, Schechter:1980gr, Schechter:1981cv}.
If right handed neutrinos exist, they could, in addition to the Dirac mass term $\overline{\nu_L} m_D \nu_R$ that is generated by the Higgs mechanism, also have a Majorana mass $\overline{\nu_R} M_M \nu_R^c$, which is forbidden for the SM fields by gauge invariance.
Here $m_D = vF$, where $v$ is the Higgs vacuum expectation value and $F$ is a matrix of Yukawa couplings between the right handed neutrinos $\nu_R$, the Higgs doublet $\phi$ and the charged lepton doublet $\ell_L = (\nu_L,\:e_L)^T$.
The scale of the eigenvalues $M_i$ of $M_M$ is entirely unknown;
different choices can have a wide range of implications
for particle physics, astrophysics and cosmology, see e.g.~\cite{Drewes:2013gca} for an overview.
Also the number $n$ of right handed neutrino states $\nu_{Ri}$ is unknown.
The minimal number $n$ required to explain the data from neutrino oscillation experiments is $n = 2$, as two light neutrinos are known to have non-zero masses.
The light neutrinos obtain their masses via a quantum mechanical mixing of the states $\nu_R$ with the left handed neutrinos $\nu_L$.
This mixing can be characterised by the entries of a matrix $\theta = m_D M_M^{-1}$.
It does not only give a small mass term $m_\nu = - m_D M_M^{-1} m_D^T = - \theta M_M \theta^T$
to the light neutrinos, but also a $\theta$-suppressed weak interaction to the heavy mass eigenstates $N_i \simeq \nu_{Ri} + \theta_{ai} \nu_{L a}^{c} + \text{c.c.}$, defined in~\eqref{HeavyMassEigenstates}.
The $N_i$ are a type of \emph{heavy neutral lepton} (HNL) with a mass $M_i$ that can be searched for experimentally.

Another motivation for the existence of the $\nu_R$ comes from cosmology.
The Yukawa couplings $F_{ai}$ generally violate CP, and the interactions of the $\nu_R$ in the early universe can potentially generate a matter-antimatter asymmetry in the primordial plasma.
At temperatures above $T_\text{sph} = \unit[130]{GeV}$~\cite{DOnofrio:2014rug} this asymmetry can be converted into a net baryon number by weak sphalerons~\cite{Kuzmin:1985mm}.
This process called \emph{leptogenesis} can either occur during the freeze out and decay of the $\nu_R$~\cite{Fukugita:1986hr} ("freeze out scenario") or during their production~\cite{Akhmedov:1998qx,Asaka:2005pn,Hambye:2016sby} ("freeze in scenario").
It is one of the most promising explanations for the \emph{baryon asymmetry of the universe} (BAU), which is believed to be the origin of baryonic matter in the present day universe, see~\cite{Canetti:2012zc} for a discussion.
The freeze in scenario is e.g.~realised in the \emph{Neutrino Minimal Standard Model} ($\nu$MSM)~\cite{Asaka:2005pn,Asaka:2005pn}.
It is particularly interesting from a phenomenological viewpoint because it is feasible for masses $M_i$ as low as \unit[10]{MeV}~\cite{Canetti:2012kh}, which are well within reach of present day experiments.

The requirement to explain the light neutrino masses imposes a lower bound on combinations of the matrix elements $\theta_{ai}$
because the differences between the eigenvalues $m_i^2$ of $m_\nu m_\nu^\dagger$ must coincide with the neutrino mass square differences obtained from neutrino oscillation experiments.
This implies that the heavy neutrinos considered here necessarily come into thermal equilibrium in the early universe~\cite{Shaposhnikov:2008pf,Hernandez:2013lza}.
To avoid tension with the observed abundances of light elements in the intergalactic medium, they should be sufficiently short lived that their decay does not disturb the primordial nucleosynthesis~\cite{Ruchayskiy:2012si}. Together with experimental constraints~\cite{Ruchayskiy:2011aa,Asaka:2011pb} this imposes a lower bound of roughly $M_i > \unit[100]{MeV}$ on those $N_i$ that participate in neutrino mass generation.%
\footnote{
This lower bound can be avoided for very feebly coupled $\nu_{Ri}$ which cannot make a significant contribution to the seesaw mechanism~\cite{Hernandez:2014fha}.
Due to their longevity (or even stability~\cite{Heeck:2015qra,Dev:2016xcp}) they could, however, be a viable Dark Matter candidates~\cite{Dodelson:1993je,Shi:1998km}, see~\cite{Adhikari:2016bei} for a review.
This is the rôle of the lightest $\nu_{Ri}$ in the $\nu$MSM~\cite{Asaka:2005an}, while the other two generate the light neutrino masses via the seesaw mechanism and the BAU via freeze in leptogenesis~\cite{Asaka:2005pn}.
}

The NA62 experiment~\cite{NA62:2017rwk} can probe precisely this mass range.
The experiment can be operated in two different modes, the \emph{kaon mode} and the \emph{beam dump mode}, as will be discussed in Section~\ref{sec:NA62}.
It is sensitive to heavy neutrinos that are produced in weak decays~\cite{Shrock:1980ct,Shrock:1981wq} of mesons or tauons~\cite{Gorbunov:2007ak} in both modes.

In the \emph{kaon mode} the SPS \unit[400]{GeV} proton beam hits a fixed target and produces a secondary positively charged hadron beam with a momentum of \unit[75]{GeV}, which comprises to $\sim \unit[6]{\%}$ of charged kaons.
Charged kaons are then used to search for the rare kaon decay $K^+ \to \pi^+ \nu \overline \nu$, which is the primary purpose of the experiment.
Among the much more frequent decays $K^+ \to \ell_a^+ \nu_a$, a small fraction ($\order{\theta^2}$) of the neutrinos appears as a heavy neutrino.
The resulting decay $K^+ \to \ell_a^+ N_i$ can be detected with a search for a peak in the missing mass distribution~\cite{Shrock:1980vy}, even if the $N_i$ itself is not seen.
Such peak searches have been performed and are known to reach deep into the region~\cite{CortinaGil:2017mqf} where leptogenesis is feasible in the $\nu$MSM~\cite{Canetti:2010aw,Canetti:2012vf,Canetti:2012kh}.
An interesting point in this context is the extremely good mass resolution, of order of few \unit{MeV}, which would allow to test the leptogenesis hypothesis~\cite{Chun:2017spz} in part of the parameter space~\cite{Hernandez:2016kel,Drewes:2016jae,Antusch:2017pkq} if any HNLs are discovered.
The disadvantage is, however, that only $N_i$ that are lighter than kaons can be produced.
In this mass range there exist strong constraints from past experiments, in particular PS191~\cite{Bernardi:1987ek}, CHARM~\cite{Bergsma:1985is}, NuTeV~\cite{Vaitaitis:1999wq,Vaitaitis:2000vc}, E949~\cite{Artamonov:2014urb}, PIENU~\cite{PIENU:2011aa}, TRIUMF-248~\cite{Britton:1992xv} and NA3~\cite{Badier:1985wg}.

In the \emph{dump mode} the target is removed and the beam is directly sent into a dump, producing large numbers of mesons and leptons of all sorts, which can decay into final states involving $N_i$.
The $N_i$ travel in the direction of the beam with a certain angular distribution, and a fraction of their decays into charged particles can be seen in the detector.
The disadvantage is that the detection relies on the $N_i$ decay, resulting in an event rate that is of $\order{\theta^4}$ in the regime where NA62 is sensitive.%
\footnote{
The number of heavy neutrinos that are produced along with a lepton of flavour $a$ can be estimated as $\sim \sigma_\nu U_a^2$, where $\sigma_\nu$ is the production cross section for light neutrinos.
Using definitions~\eqref{Udef}, the number of events with lepton flavour $b$ in the final state that can be seen in a detector can then be estimated as
$\sim \left(1 - e^{\Gamma L} \right) \sigma_\nu U_{ai}^2 U_{bi}^2 / U_i^2$, where $\Gamma\propto \frac{G_F^2}{96 \pi^3} U_i^2 M_i^5$ is the HNL decay rate (see~\cite{Bondarenko:2018ptm} for recent more precise computations) and $L$ is the length of the detector.
For $L\Gamma \ll 1$ one can expand the exponential and the event rate is of order $\order{\theta^4}$.
}
On the other hand, one is not restricted to $N_i$ below the kaon mass because the $N_i$ can be produced in the decay of any meson.
This in principle extends the range of $M_i$ that can be probed up to the $B$-meson mass,%
\footnote{The IceCube experiment can cover a similar mass range as NA62 with searches for double cascade events~\cite{Coloma:2017ppo}, though with a somewhat lower sensitivity.
A significant improvement in the entire mass range below the $B$-meson mass could be achieved with the proposed SHiP experiment~\cite{Graverini:2214085}.
Part of this mass range could also be probed by a FASER~\cite{Kling:2018wct}, CODEX-b~\cite{Gligorov:2017nwh} or a MATHUSLA-type~\cite{Chou:2016lxi,Curtin:2018mvb} upgrade to the LHC in the future.
Above the $B$-meson mass, displaced vertex searches at high energy hadron~\cite{Helo:2013esa,Izaguirre:2015pga,Antusch:2016vyf,Gago:2015vma,Antusch:2017hhu} or lepton~\cite{Antusch:2016vyf,Antusch:2017pkq} colliders would be more sensitive, cf.~\cite{Antusch:2016ejd} for a summary.
Neutrinos that are heavy enough to decay promptly can leave distinct lepton number and flavour violating signatures in high energy collisions, see~\cite{Deppisch:2015qwa, Cai:2017mow} for a recent review.}
though the number of produced $B$-mesons is so small that the $D$-meson mass is practically the upper limit for the range of $M_i$ that can be probed with NA62.
In this paper we study the sensitivity of the NA62 experiment when operated in dump mode to heavy neutrinos, and how it depends on assumptions about the heavy neutrino flavour mixing patter in view of the most recent neutrino oscillation data.

\section{Heavy Neutrino Flavour mixing}

\subsection{Heavy Neutrinos in the type-I Seesaw Model}

The most general renormalisable addition to the SM Lagrangian that can be constructed from SM fields and the $\nu_R$ reads
\begin{equation}
    \mathcal L_{\nu_R}
  = \iu \overline{\nu_R}_i \slashed\partial \nu_{Ri}
  - \frac{1}{2} \left( \overline{\nu_R^c}_i(M_M)_{ij}\nu_{Rj}
    + \overline{\nu_R}_i(M_M^\dagger)_{ij}\nu_{}^c \right)
  - F_{ai} \overline \ell_a \varepsilon \phi \nu_{Ri}
  - F_{ai}^* \overline{\nu_R}_i \phi^\dagger \varepsilon^\dagger \ell_a
\ . \label{eq:Lagrangian}
\end{equation}
Here $\varepsilon$ is the antisymmetric SU(2) tensor and we have suppressed all SU(2) indices.
We work in this minimal model, which adds no other New Physics to the SM.
This is literally the case in the $\nu$MSM, to which our results can directly be applied.
It is also valid in models where all other New Physics
a) involves only new particles that are much heavier than \unit[27]{GeV} (the centre of mass collision energy of the NA62 experiment) and
b) does not significantly contribute to neutrino mass generation.
After electroweak symmetry breaking the Higgs field obtains an expectation value $v = \unit[174]{GeV}$, which generates the Dirac mass term $\overline{\nu_L}_a (m_D)_{a i}\nu_{Ri}$ from the term $F_{ai}\overline \ell_a \varepsilon \phi \nu_{Ri}$.
Due to the seesaw hierarchy $M_M \gg m_D$ we expand all quantities to second oder in the $\theta_{a i}$ in what follows.
The light and heavy mass eigenstates after electroweak symmetry breaking are then
\begin{align}
    \upnu_i &
  = \left[ V_\nu^{\dagger}\nu_L-U_\nu^{\dagger}\theta\nu_R^c + V_\nu^{T}\nu_L^c-U_\nu^{T}\theta\nu_R \right]_i
\ , 
  & N_i &
  = \left[V_N^\dagger\nu_R+\Theta^{T}\nu_L^{c} + V_N^T\nu_R^c+\Theta^{\dagger}\nu_L\right]_i
\ , \label{HeavyMassEigenstates}
\end{align}
respectively.
Here $V_\nu = \left(1 - \frac{1}{2}\theta\theta^\dagger \right) U_\nu$ is the light neutrino mixing matrix that diagonalises the light neutrino mass matrix
\begin{equation}
    m_\nu
  = - m_D M_M^{-1} m_D^T
  = - \theta M_M \theta^T
\label{seesaw}
\end{equation}
and $U_\nu$ is its unitary part, known as Pontecorvo-Maki-Nakagawa-Sakata matrix.
$V_N = \left(1 - \frac{1}{2} \theta^T \theta^*\right) U_N$ is the equivalent matrix that diagonalises the heavy neutrino mass matrix $M_N = M + \frac{1}{2} \left(\theta^\dagger \theta M + M^T \theta^T \theta^{*}\right)$ after electroweak symmetry breaking, and $\Theta = \theta U_N^*$.
In practice we can set $V_N\simeq 1$ and $\Theta\simeq \theta$.
The mixing gives a $\theta$-suppressed weak interaction to the heavy mass eigenstates~\eqref{HeavyMassEigenstates}.
The coupling of the mass eigenstates $N_i$ to the SM can then be described by the Lagrangian contribution
\begin{align}
    \mathcal L
  \supset&
  - \frac{g}{\sqrt{2}}\overline N_i \Theta^\dagger_{i \alpha}\gamma^\mu e_{L \alpha} W^+_\mu
  - \frac{g}{\sqrt{2}}\overline{e_L}_\alpha\gamma^\mu \Theta_{\alpha i} N_i W^-_\mu\notag\\
 &- \frac{g}{2\cos\theta_W}\overline N_i \Theta^\dagger_{i \alpha}\gamma^\mu \nu_{L \alpha} Z_\mu
  - \frac{g}{2\cos\theta_W}\overline{\nu_L}_\alpha\gamma^\mu \Theta_{\alpha i} N_i Z_\mu\notag\\
 &- \frac{g}{\sqrt{2}}\frac{M_i}{m_W}\Theta_{\alpha i} h \overline{\nu_L}_\alpha N_i
  - \frac{g}{\sqrt{2}}\frac{M_i}{m_W}\Theta^\dagger_{i \alpha} h \overline N_i \nu_{L\alpha}
\ . \label{WeakWW}
\end{align}
The first two lines are the couplings of the $N_i$ to the weak currents due to the mixing~$\Theta$, and the last line is the Yukawa coupling to the physical Higgs field $h$ in the unitary gauge expressed using the definition of $\theta$ and the relation $m_W = \frac{1}{2}v g$, with $g$ the weak gauge coupling constant.
This Yukawa term is not relevant for NA62 because the branching ratios for decays mediated by virtual Higgs bosons are suppressed due to the small Yukawa couplings of the kinematically accessible electron and muon final states.
Due to the interactions~\eqref{WeakWW} the heavy neutrinos can participate in all processes that involve ordinary neutrinos if this is kinematically allowed, but with amplitudes suppressed by~$\Theta_{ai}$.
The event rates are proportional to combinations of the quantities
\begin{equation}
U_{ai}^2 = \abs{\Theta_{ai}}^2
\ .
\end{equation}
In the \emph{symmetry protected scenario} defined further below the heavy neutrinos tend to have quasi-degenerate masses.
If the mass splitting is smaller than the experimental resolution, it is not possible to resolve them individually.
It is convenient to introduce the notation
\begin{align}\label{Udef}
    U_a^2
 &= \sum_i U_{ai}^2
\ ,
 &  U_i^2
 &= \sum_a U_{ai}^2
\ ,
 &  U^2
 &= \sum_i U_i^2
\ .
\end{align}

If the $\nu_{Ri}$ are the sole origin of the light neutrino masses, then the properties of the $N_i$
are strongly constrained by relation~\eqref{seesaw}.
This connection has been studied by various authors, see e.g.~\cite{Gorbunov:2007ak,Kersten:2007vk,Shaposhnikov:2008pf,Gavela:2009cd,Ruchayskiy:2011aa,Asaka:2011pb,Gorbunov:2013dta,Hernandez:2016kel,Drewes:2016jae}.
The constraints depend on the number $n$ of heavy neutrinos.
In general, the number of parameters in the seesaw Lagrangian~\eqref{eq:Lagrangian} in addition to those in the SM is $7n-3$.
$n$ of them can be identified with the masses $M_i$, the remaining ones are mixing angles and phases.
We discuss the cases $n = 2$ and $n = 3$ below.
The $n = 2$ scenario is the minimal model that allows to explain the two observed neutrino mass differences and predicts the lightest neutrino to be massless ($m_\text{lightest} = 0$).
It also effectively describes the neutrino mass generation in the $\nu$MSM, which in principle contains three heavy neutrinos.
The lightest $N_i$ in the $\nu$MSM is a Dark Matter candidate, and the strong observational constraints on its interactions imply that it has no significant effect on the light neutrino masses, see e.g.~the review~\cite{Adhikari:2016bei} and references therein.
The $n = 3$ scenario is the minimal model consistent with neutrino oscillation data if $m_\text{lightest} > 0$; in many models with extended gauge sectors it is also motivated by the requirement of anomaly freedom and the fact that there are three generations in the SM.

\subsection{Connection to light neutrino oscillation data}

\begin{table}
\centering
\begin{tabular}{l r@{}l r@{}l r@{}l}
    \toprule
    \multicolumn{3}{c}{Variables}
  & \multicolumn{2}{c}{NO}
  & \multicolumn{2}{c}{IO}
 \\ \midrule
    \multirow{3}{*}{Masses}
  & $m$&$_1^2$
  & $m$&$_\text{lightest}^2$
  & $m$&$_\text{lightest}^2 - \Delta m^2_{32}- \Delta m^2_\text{sol}$
 \\
  & $m$&$_2^2$
  & $m$&$_\text{lightest}^2 + \Delta m^2_\text{sol}$
  & $m$&$_\text{lightest}^2 - \Delta m^2_{32}$
 \\
  & $m$&$_3^2$
  & $m$&$_\text{lightest}^2 + \Delta m^2_{31}$
  & $m$&$_\text{lightest}^2$
 \\ \midrule
    \multirow{2}{*}[-.5ex]{Differences}
  & larger $\Delta m$&$^2$
  & $\Delta m$&$^2_{31} = m_3^2 - m_1^2$
  & $\Delta m$&$^2_{32} = m_3^2 - m_2^2$
 \\ \cmidrule{2-7}
  & ($n=2$) $\Delta m$&$^2_\text{atm}$
  & $m$&$_3^2$
  & $m$&$_1^2 \equiv m_2^2 + \order{\Delta m^2_\text{sol}/\Delta m^2_\text{atm}}$
 \\ \bottomrule
\end{tabular}
\caption{%
Definitions of the neutrino masses and their differences, for "normal ordering" (NO) and "inverted ordering" (IO).
The smaller mass difference ("solar mass difference") is given by $\Delta m^2_\text{sol} = m_2^2 - m_1^2$ for both hierarchies, while the larger mass difference is hierarchy depended.
For historical reasons one also defines the "atmospheric mass difference" $\Delta m^2_\text{atm} = \abs{m_3^2 - m_1^2}$.
One can roughly identify $\Delta m^2_\text{atm}$ and $\Delta m^2_\text{sol}$ as the larger and smaller mass splitting, respectively.
The values for $m^2_\text{atm}$ differ slightly for the two hierarchies and the difference is order $\Delta m^2_\text{sol}/\Delta m^2_\text{atm}$.
For $n = 2$ the lightest neutrino is massless ($m_\text{lightest} = 0$), which allows in both cases to identify $\Delta m^2_\text{atm}$ with one of the neutrino masses.
}
\label{tab:definitions}
\end{table}

The smallness of the light neutrino masses $m_i$ can be explained by the seesaw relation~\eqref{seesaw} in different ways.
One possibility is that the $N_i$ are superheavy.
In this case the smallness of the $m_i$ is due to the smallness of $v/M_1$, where $M_1$ is the smallest eigenvalue of $M_M$.
This conventional seesaw mechanism cannot work for the masses $M_i<v$ that are accessible to NA62.
If one simply "downscales" the $M_i$ to experimentally accessible values, one would roughly expect $\abs{F_{ai}}^2 \sim M_i m_a / v^2$ and $U_{ai}^2\sim m_a/M_i$,
where the symbol "$\sim$" indicates that we have neglected light neutrino mixing, i.e., we neglect the difference between mass and flavour eigenstates and assume that the flavour eigenstate $\nu_{L a}$  has the mass $m_a$ of the mass eigenstate that it is mostly composed of.
If we would further assume that all $N_i$ have the same mass $\overline M$, then we can define the parameters
\begin{align}\label{F0}
    F_0^2
 &= \frac{\overline M}{v^2} \sqrt{\Delta m_\text{atm}^2 + m_\text{lightest}^2}
 & \text{and} &
 & U_0^2
 &= \frac{1}{\overline M} \sqrt{\Delta m_\text{atm}^2 + m_\text{lightest}^2}
\end{align}
to quantify the deviation of the Yukawa couplings and mixings from the "naive seesaw expectation".
Here $\Delta m_\text{atm}^2$ is the larger of the two observed neutrino mass splittings, with $\Delta m_\text{sol}^2$ being the smaller one, cf.~Table~\ref{tab:definitions}.
This would be very discouraging for experimentalists because the production rates for HNLs with $U^2 \simeq U_0^2$ is tiny even if the \emph{seesaw scale} $M_1$ is as low as \unit[100]{MeV}.

Another possibility is that the $m_i$ are small as the result of a slightly broken symmetry.
Comparably low values of the $M_i$ are technically natural because the $B - L$ symmetry of the SM (which is in general broken by the $M_i$) is restored in the limit of all $M_i \to 0$.
An approximate $B - L$ symmetry can be realised even if $M_i\neq0$ if the $\nu_{Ri}$ come in pairs with equal mass $M_i = M_j$ and couplings $F_{ai} = i F_{a j}$ that can be represented by Dirac spinors $\nu_{Ri} + \nu_{Rj}^c$~\cite{Shaposhnikov:2006nn,Kersten:2007vk,Moffat:2017feq}.
This \emph{symmetry protected scenario} provides a theoretical motivation for a low scale seesaw and allows for experimentally accessible $U_{ai}^2 \gg U_0^2$.
Specific examples that motivate this limit include in "inverse seesaw" type scenarios~\cite{Wyler:1982dd,Mohapatra:1986aw,Mohapatra:1986bd,Bernabeu:1987gr}, a "linear seesaw"~\cite{Akhmedov:1995ip,Akhmedov:1995vm}, scale invariant models~\cite{Khoze:2013oga}, some technicolour-type models~\cite{Appelquist:2002me,Appelquist:2003uu} or the $\nu$MSM~\cite{Shaposhnikov:2006nn}.

\begin{table}
\centering
\begin{tabular}{l r@{}l r@{}l r@{}l}
    \toprule
    \multicolumn{3}{c}{Variables}
  & \multicolumn{2}{c}{NO}
  & \multicolumn{2}{c}{IO}
 \\ \midrule
    \multirow{2}{*}{Differences}
  & (smaller) $\Delta m$&$^2_\text{sol}$
  & $7$&$.40\phantom{0}\times \unit[10^{-5}]{eV^2}$
  & $7$&$.40\phantom{0}\times \unit[10^{-5}]{eV^2}$
 \\
  & larger $\Delta m$&$^2$
  & $2$&$.515\times \unit[10^{-3}]{eV^2}$
  & $- 2$&$.483\times \unit[10^{-3}]{eV^2}$
 \\ \midrule
    \multirow{3}{*}{Angles}
  & $\sin$&$^2\uptheta_{12}$
  & $0$&$.307$
  & $0$&$.307$
 \\
  & $\sin$&$^2\uptheta_{13}$
  & $0$&$.02195$
  & $0$&$.02212$
 \\
  & $\sin$&$^2\uptheta_{23}$
  & $0$&$.565$
  & $0$&$.572$
 \\ \bottomrule
\end{tabular}
\caption{%
Best fit values of neutrino mass differences and mixing angles from the NuFIT~3.1 release by the $\nu$-fit collaboration~\cite{Esteban:2016qun, nufit}, for "normal ordering" (NO) and "inverted ordering" (IO).
The mass differences are defined in Table~\ref{tab:definitions}.
}
\label{tab:active_bounds}
\end{table}

A convenient way to connect the heavy neutrino couplings to neutrino oscillation data is provided by the Casas-Ibarra parametrisation~\cite{Casas:2001sr}
\begin{align}\label{CasasIbarraDef}
    F
 &= \frac{1}{v} U_\nu \sqrt{m_\nu^\text{diag}} \mathcal R \sqrt{M_M^\text{diag}}
\ ,
 &  \theta
 &= U_\nu \sqrt{m_\nu^\text{diag}} \mathcal R \sqrt{M_M^\text{diag}}^{-1}
\ .
\end{align}
Here $(m_\nu^\text{diag})_{ij} = \delta_{ij} m_i$ and $(M_M^\text{diag})_{ij} = \delta_{ij} M_i$.
We use the common parameterisation
\begin{equation}
    U_\nu
  = V^{(23)} U_\delta V^{(13)} U_{-\delta} V^{(12)} \diag(e^{\iu \alpha_1/2},\: e^{\iu \alpha_2 /2},\: 1)
\ , \label{PMNS}
\end{equation}
with $U_{\pm \delta} = \diag(1,\: e^{\mp \iu \delta /2},\: e^{\pm \iu \delta /2})$, and non-vanishing entries of $V^{(ab)}$ for $a = e,\:\mu,\:\tau$ are
\begin{align}
    V^{(ab)}_{aa}
 &= V^{(ab)}_{bb}
  = \cos \uptheta_{ab}
\ ,
 &  V^{(ab)}_{ab}
 &= -V^{(ab)}_{ba}
  = \sin \uptheta_{ab}
\ ,
 &  V^{(ab)}_{cc}
 &= 1
 && \text{for}
 &  c
 &  \neq a,\:b
 \ .
\end{align}
Here $\uptheta_{ab}$ are the light neutrino mixing angles.
For the purpose of fixing the parameters in $U_\nu$, we approximate $V_\nu\simeq U_\nu$ and use the data set NuFIT~3.1 from the $\nu$-fit collaboration~\cite{Esteban:2016qun, nufit}, cf.~Table~\ref{tab:active_bounds}.
The NuFIT~3.2 update released shortly after our work was finalised.
We checked that our results (in particular the benchmark scenarios listed in Table~\ref{tab:benchmark scenarios}) are not significantly affected by the update, cf.~Figures~\ref{fig:Chi2NO} and~\ref{fig:Chi2IO}.

In addition to the neutrino oscillation data, $N_i$ with masses below the electroweak scale are constrained by a number of direct and indirect searches, see e.g.~\cite{Gorbunov:2014ypa,Drewes:2015iva} and references therein.
With the exception of neutrinoless double $\beta$ decay~\cite{Bezrukov:2005mx,Asaka:2011pb,LopezPavon:2012zg,Asaka:2013jfa,Lopez-Pavon:2015cga,Drewes:2016lqo,Hernandez:2016kel}, the indirect observables are mostly irrelevant in the mass range that can be tested by NA62.

\subsection{The Minimal Model with $n = 2$}\label{sec:nis2}

For $n = 2$ the matrix $\mathcal{R}$ in the parameterisation~\eqref{CasasIbarraDef} can be expressed in terms of a single complex parameter $\omega$,
\begin{align}
    \mathcal{R}^\text{NO}
 &= \begin{pmatrix}
        0 & 0
     \\ \cos \omega & \sin \omega
     \\ -\xi \sin \omega & \xi \cos \omega
    \end{pmatrix}
\ ,
 &  \mathcal{R}^\text{IO}
 &= \begin{pmatrix}
        \cos \omega & \sin \omega
     \\ -\xi \sin \omega & \xi \cos \omega
     \\ 0 & 0
    \end{pmatrix}
\ ,
\end{align}
where $\xi = \pm 1$ and "NO" and "IO" refer to "normal ordering" and "inverted ordering" amongst the light neutrino masses $m_i$.
Moreover, only one combination of the Majorana phases is physical.
For normal ordering this is $\alpha_2$, while for inverted ordering it is $\alpha_2 - \alpha_1$.
We can therefore without loss of generality set $\alpha_2 \equiv \alpha$ and $\alpha_1 = 0$.
The symmetry protected regime in this parametrisation (and for $\xi = 1$) corresponds to small values of the symmetry breaking parameters
\begin{align}
    \upepsilon
  &\equiv
    e^{-2 \Im \omega }
\ ,
  & \upmu
 &= \frac{\Delta M}{\overline M}
\label{B-L_parameters}
&&\text{with}
   & \Delta M
 &= \frac{M_2 - M_1}{2}
\ ,
  &   \overline M
  &= \frac{M_2 + M_1}{2}
\ .
\end{align}
In terms of these parameters one can express the total interaction strength $U^2$ of the heavy neutrinos as
\begin{subequations}
\begin{equation}
    U^2
  = \frac{1}{1 - \upmu^2} \left[ 2 \upmu \cos(2 \Re \omega) \frac{\Delta m}{\overline M}
  + \left(\upepsilon + \frac{1}{\upepsilon} \right) \frac{\overline m}{\overline M} \right]
\ ,
\end{equation}
where
\begin{align}
    \Delta m
 &= \frac12
    \begin{cases}
        m_2 - m_3 & \text{for NO}
     \\ m_1 - m_2 & \text{for IO}
    \end{cases}
\ ,
  & \overline m
 &= \frac12
    \begin{cases}
        m_2 + m_3 & \text{for NO}
     \\ m_1 + m_2 & \text{for IO}
    \end{cases}
\ .
\end{align}
\end{subequations}
While $\upmu$ can in principle be set exactly to zero, there are theoretical lower limits on $\upepsilon$ from the requirements to remain perturbative ($\abs{F_{ai}} \ll 4\pi$) and justify the expansion in $\theta$ ($U^2\ll 1$).
Practically the lower limit on $\upepsilon$ from the existing experimental constraints in the mass range accessible to NA62 is stronger than these theoretical considerations.
The relations between the individual $U_{ai}^2$ and the parameters in $U_\nu$ are given in Appendix~\ref{App:Mixing}.

\begin{figure}
\begin{subfigure}{0.5\linewidth}
\includegraphics[width=\linewidth]{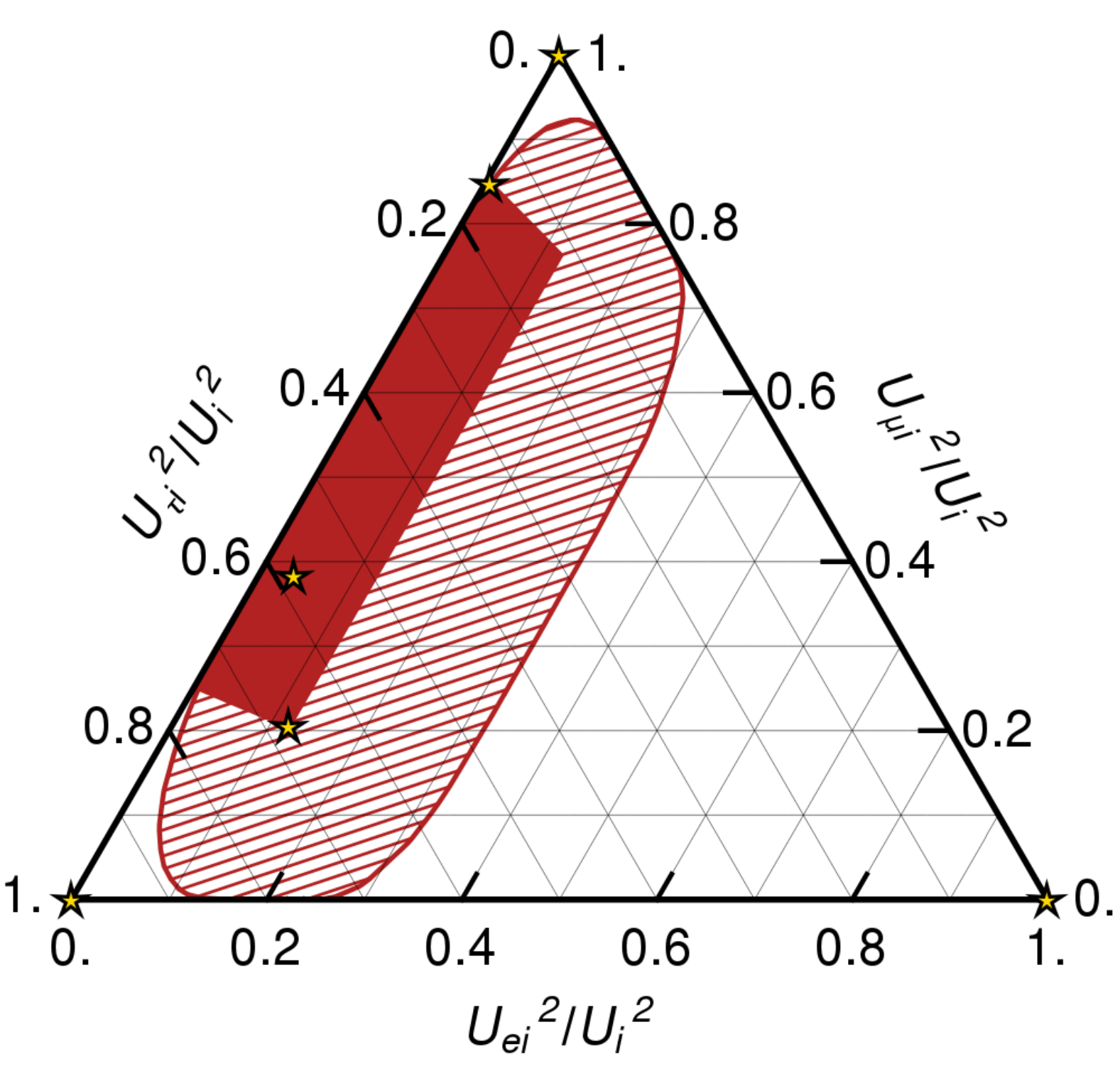}
\caption{Normal ordering.}
\label{fig:allowed areas normal ordering}
\end{subfigure}
\begin{subfigure}{0.5\linewidth}
\includegraphics[width=\linewidth]{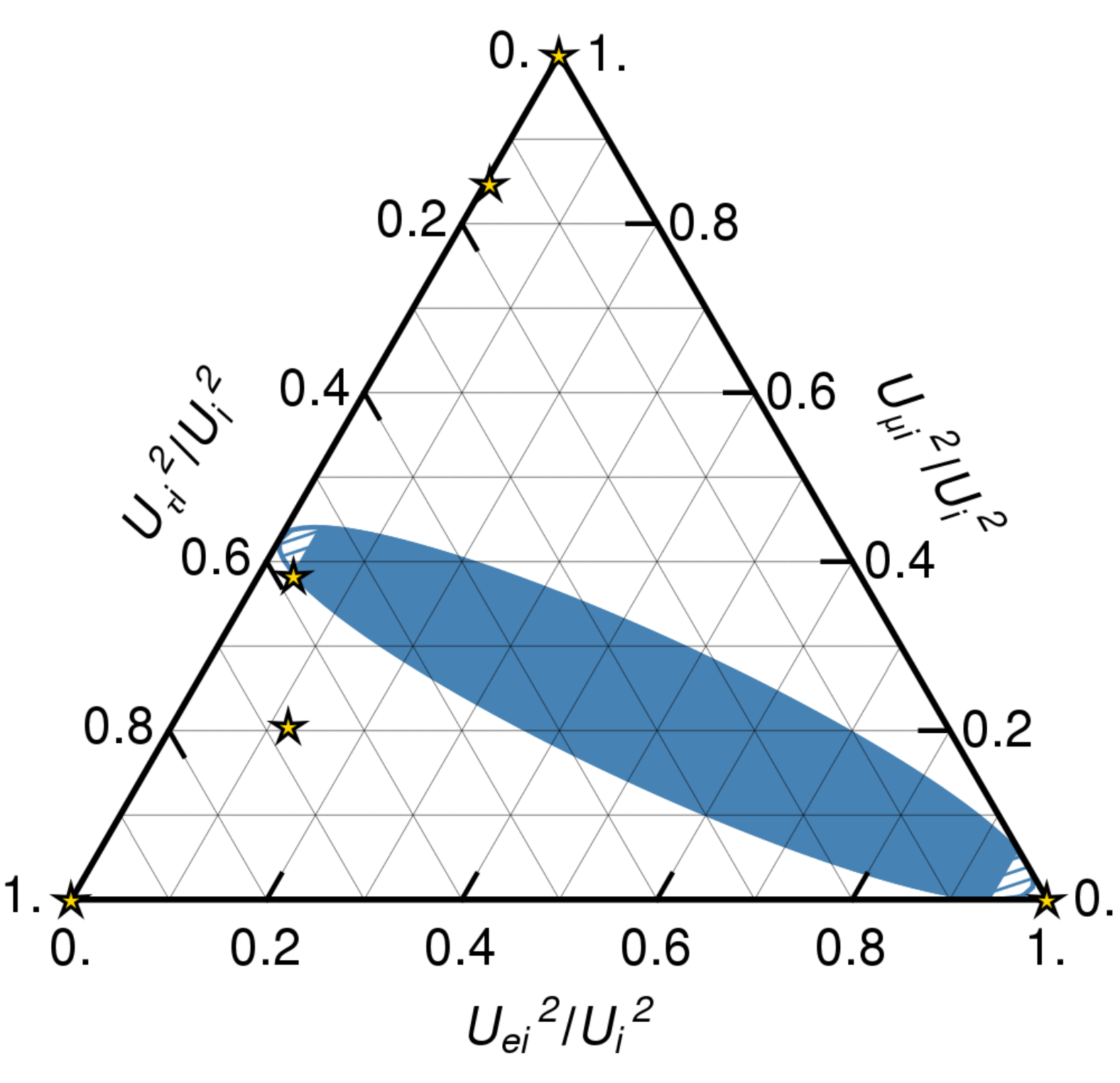}
\caption{Inverted ordering.}
\label{fig:allowed areas inverted ordering}
\end{subfigure}
\caption{Allowed range of $U_a^2/U^2$ in the $n = 2$ model for arbitrary parameter choices (hashed region) and in the symmetric limit (filled region) for normal ordering in Panel~(\subref{fig:symmetric limit normal ordering}) and inverted ordering in Panel~(\subref{fig:symmetric limit inverted ordering}) of light neutrino masses.
For an experiment with the sensitivity of NA62, the minimal model with $n = 2$ predicts the $U_{ai}^2/U_i^2$ to lie within the filled areas.
Mixing patterns in the extended hashed regions can only be made consistent with light neutrino oscillation data for total $U_i^2 \times M_i/\unit{GeV} < 10^{-11}$, cf.~Figure~\ref{fig:SymmLimApplicability}.
The stars mark the benchmark scenarios given in Table~\ref{tab:our benchmark scenarios}.}
\label{fig:AllowedAreas}
\end{figure}

\begin{figure}
\begin{subfigure}{0.5\linewidth}
\includegraphics[width=\linewidth]{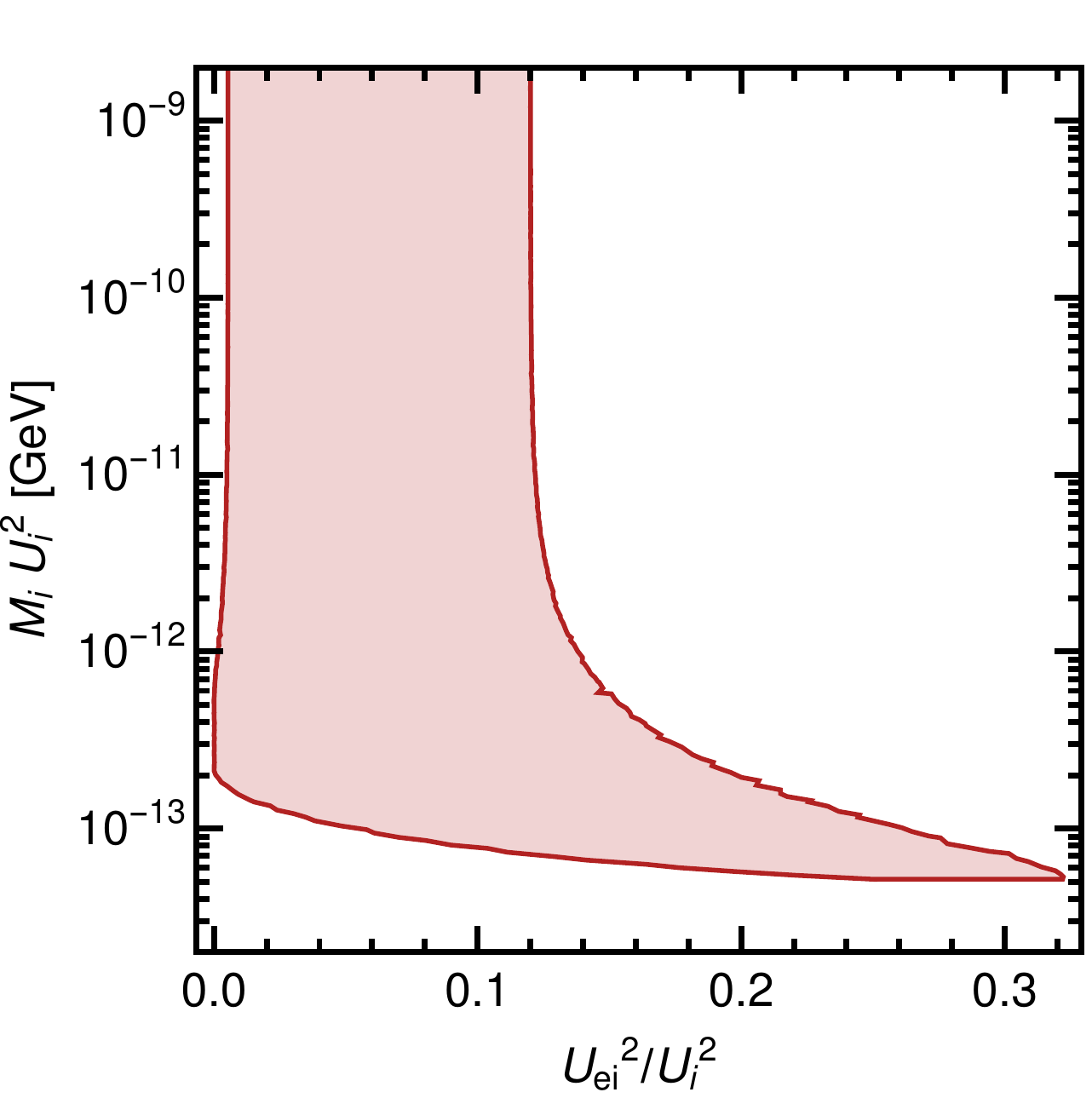}
\caption{Normal ordering.}
\label{fig:symmetric limit normal ordering}
\end{subfigure}
\begin{subfigure}{0.5\linewidth}
\includegraphics[width=\linewidth]{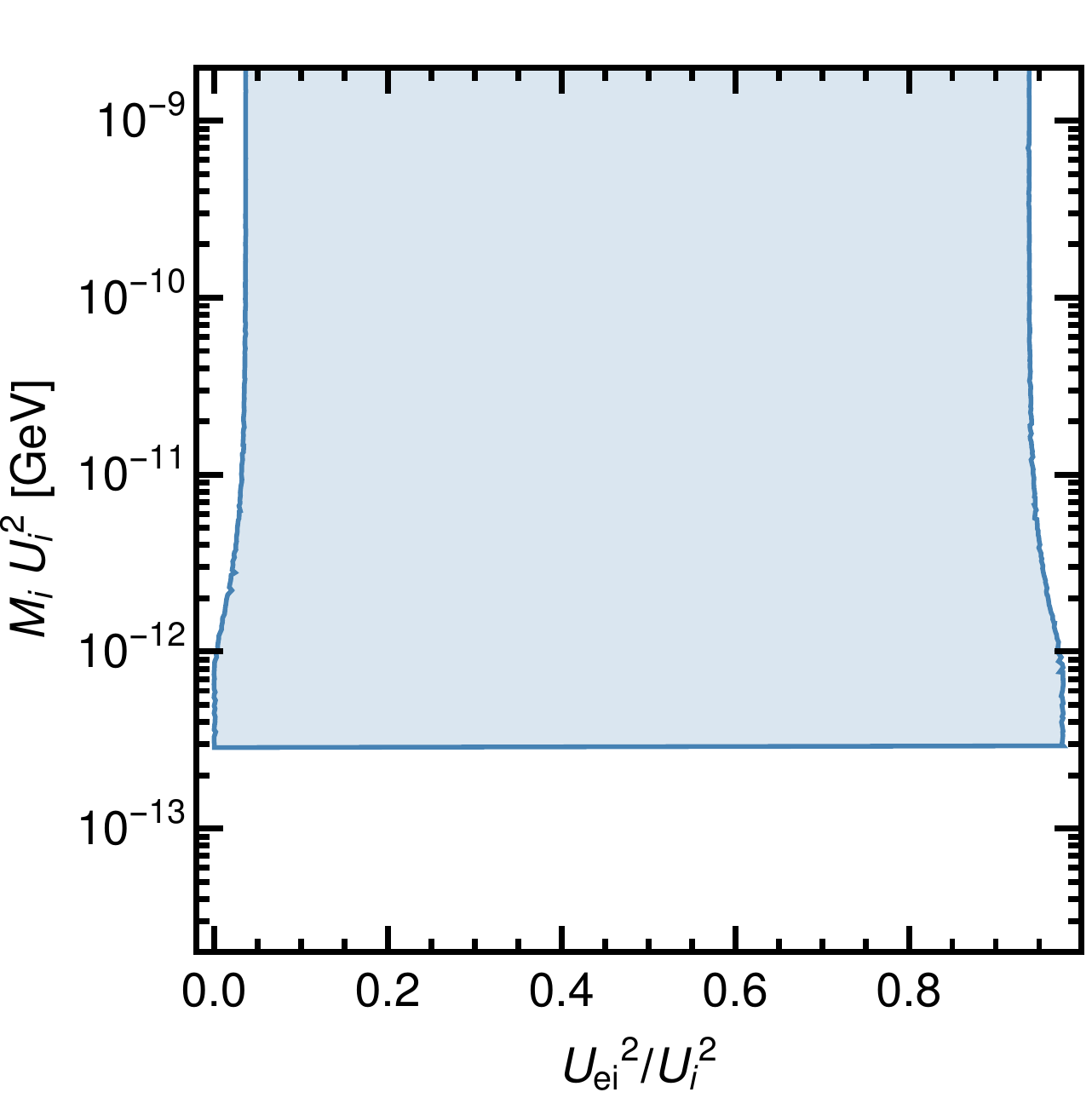}
\caption{Inverted ordering.}
\label{fig:symmetric limit inverted ordering}
\end{subfigure}
\caption{The applicability of the symmetric limit can be illustrated by plotting the allowed range of $U_{e i}^2/U_i^2$ for different values of $U_i^2 M_i$, which is in good approximation independent of $M_i$.
For both, normal ordering shown in Panel~(\subref{fig:symmetric limit normal ordering}) and inverted ordering shown in Panel~(\subref{fig:symmetric limit inverted ordering}), the allowed range of $U_{e i}^2/U_i^2$ is independent of $U_i^2 M_i$ for values of $U_i^2 > 10^{-11} \frac{\unit{GeV}}{M_i}$.
These regions correspond to the interval predicted in the symmetric limit.
}
\label{fig:SymmLimApplicability}
\end{figure}

We define the \emph{symmetric limit} by setting $\upmu = 0$ (i.e.~$M_1 = M_2 = \overline M$) and keeping only terms of order $1/\upepsilon$.
Values $U_{ai}^2 \gg U_0^2$ that yield sizeable branching fractions from and to SM particles for heavy neutrinos necessarily require $\upepsilon \ll 1$.%
\footnote{Alternatively we could consider the case $1/\upepsilon \ll 1$.
Both cases can be related by the observation that swapping the signs of $\xi$, $\Im \omega$ and $\Delta M$ as well as and changing $\Re \omega\to\pi-\Re \omega$ swaps the labels of $N_1$ and $N_2$, with no physical consequences.}
This requires cancellations in the light neutrino mixing matrix~\eqref{seesaw}, which should be considered "fine tuned" unless one also takes the limit $\upmu\ll1$, in which case a generalisation of the global $B-L$ symmetry in the SM is approximately respected by the Lagrangian~\eqref{eq:Lagrangian}.
Hence, the condition $\upmu\ll 1$ that is required for leptogenesis with $n = 2$~\cite{Asaka:2005pn} is automatically fulfilled in a (technically) natural way in the symmetric limit.
The limit $\upmu,\: \upepsilon \ll 1$ yields
\begin{align}\label{SymmLimitCouplings}
    M_1
 &= M_2
  = \overline M
\ ,
 &  U_{a 1}^2
 &= U_{a 1}^2
  = \frac12 U_a^2
\ .
\end{align}
A remarkable feature of the limit $\upepsilon \rightarrow 0$ is that the ratios $U_a^2/U^2$ become independent of the seesaw scale $\overline M$ and the unknown parameter $\Re \omega$ irrespective of the choice of $\upmu$.
That is, they can be expressed in terms of the low energy parameters in $U_\nu$ alone.
On one hand, this in principle allows one to determine the Majorana phase $\alpha$ from measurements of the $U_a^2$~\cite{Caputo:2016ojx}.
On the other hand this means that existing neutrino oscillation data can be used to make predictions for the allowed range of $U_a^2/U^2$ in the minimal seesaw model.
A simple estimate can be made if one fixes the light neutrino mass splittings and mixing angles to the best fit values given in Table~\ref{tab:active_bounds} and freely varies all other parameters, see Figure~\ref{fig:AllowedAreas}.
While in principle the entire hashed regions are allowed, NA62 can only find $N_i$ with mixing angles which are large enough that the symmetry protected limit can be applied, cf.~Figure~\ref{fig:SymmLimApplicability}.
Hence, if NA62 finds HNLs with $U_a^2/U^2$ outside the filled regions, this would clearly disfavour the minimal model with $n = 2$.

The present status of neutrino oscillation experiments allows to do a more quantitative analysis.
One can use the statistical information about the light neutrino parameters gathered in various neutrino oscillation experiments to obtain a probability distribution for the $U_a^2/U^2$.
As input we use the results of the NuFIT~3.1 release for this purpose, from which one can first determine likelihoods for $\Delta m^2_{31}$, $\Delta m^2_{32}$ and all parameters in $U_\nu$ with the exception of the Majorana phase $\alpha$.
Due to the large deviations from the Gaussian limit, for $\delta$ and $\uptheta_{23}$ we use the two-dimensional $\Delta \chi^2$ projection, while for all other parameters we use the one-dimensional projections.
The total $\Delta \chi^2$ for a set of parameters is then given by the sum of the individual $\Delta \chi^2$.

\begin{figure}
\begin{subfigure}{0.5\linewidth}
\includegraphics[width=\linewidth]{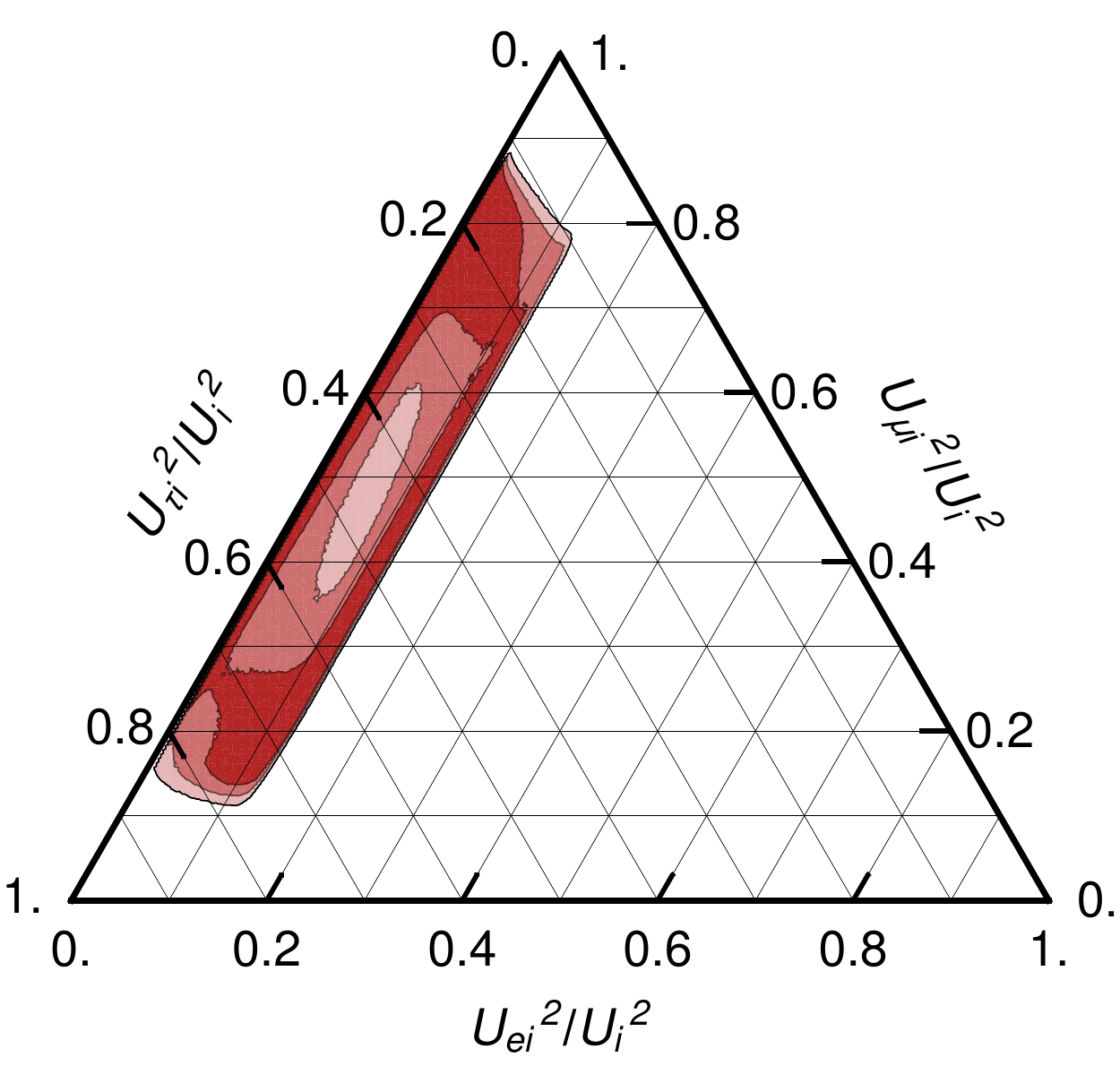}
\caption{Flat prior on $\alpha$.}
\label{fig:NOalpha}
\end{subfigure}
\begin{subfigure}{0.5\linewidth}
\includegraphics[width=\linewidth]{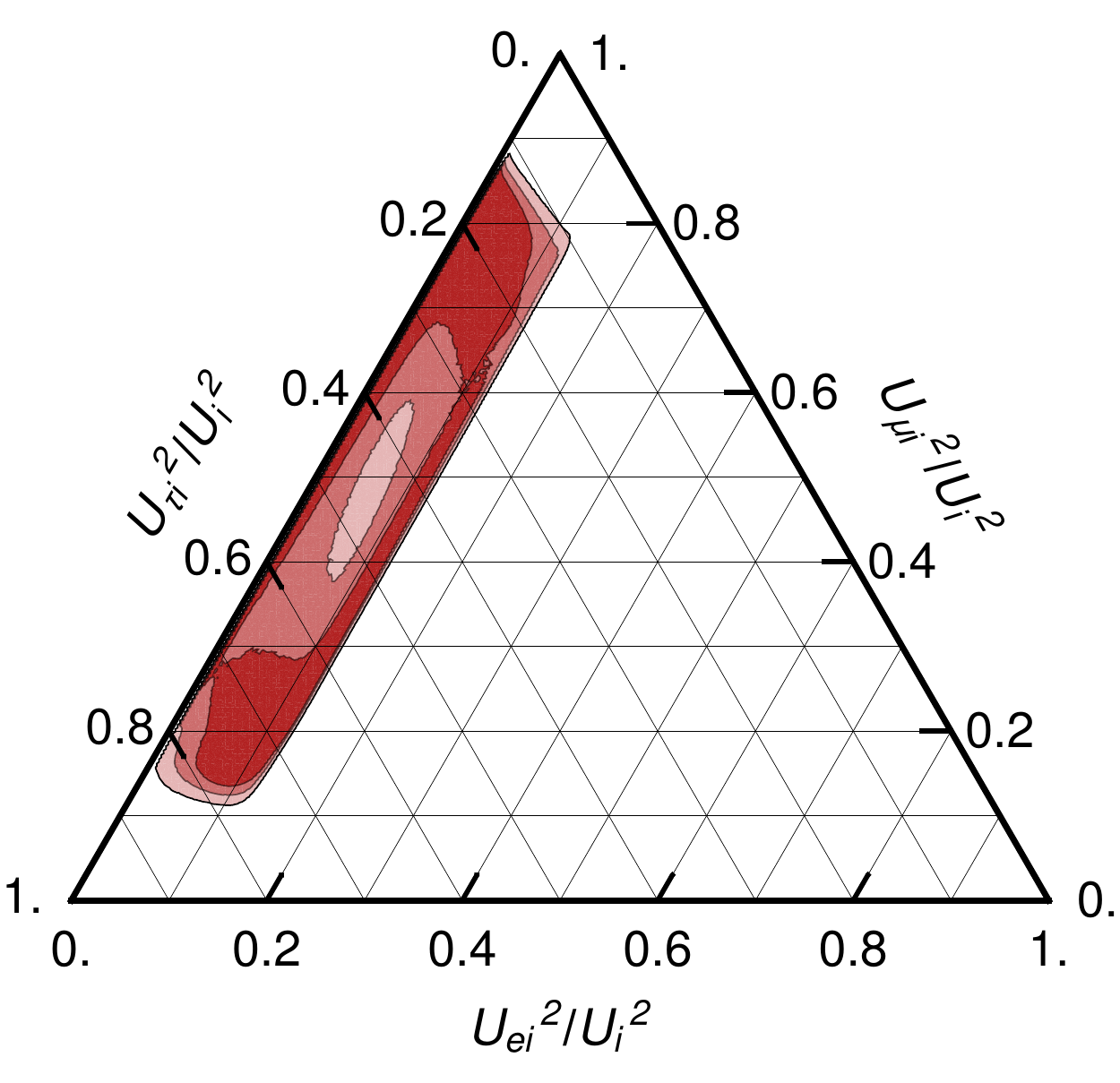}
\caption{Flat prior on $\sin(\alpha/2+\delta)$.}
\label{fig:NOsin}
\end{subfigure}\\
\begin{subfigure}{0.5\linewidth}
\includegraphics[width=\linewidth]{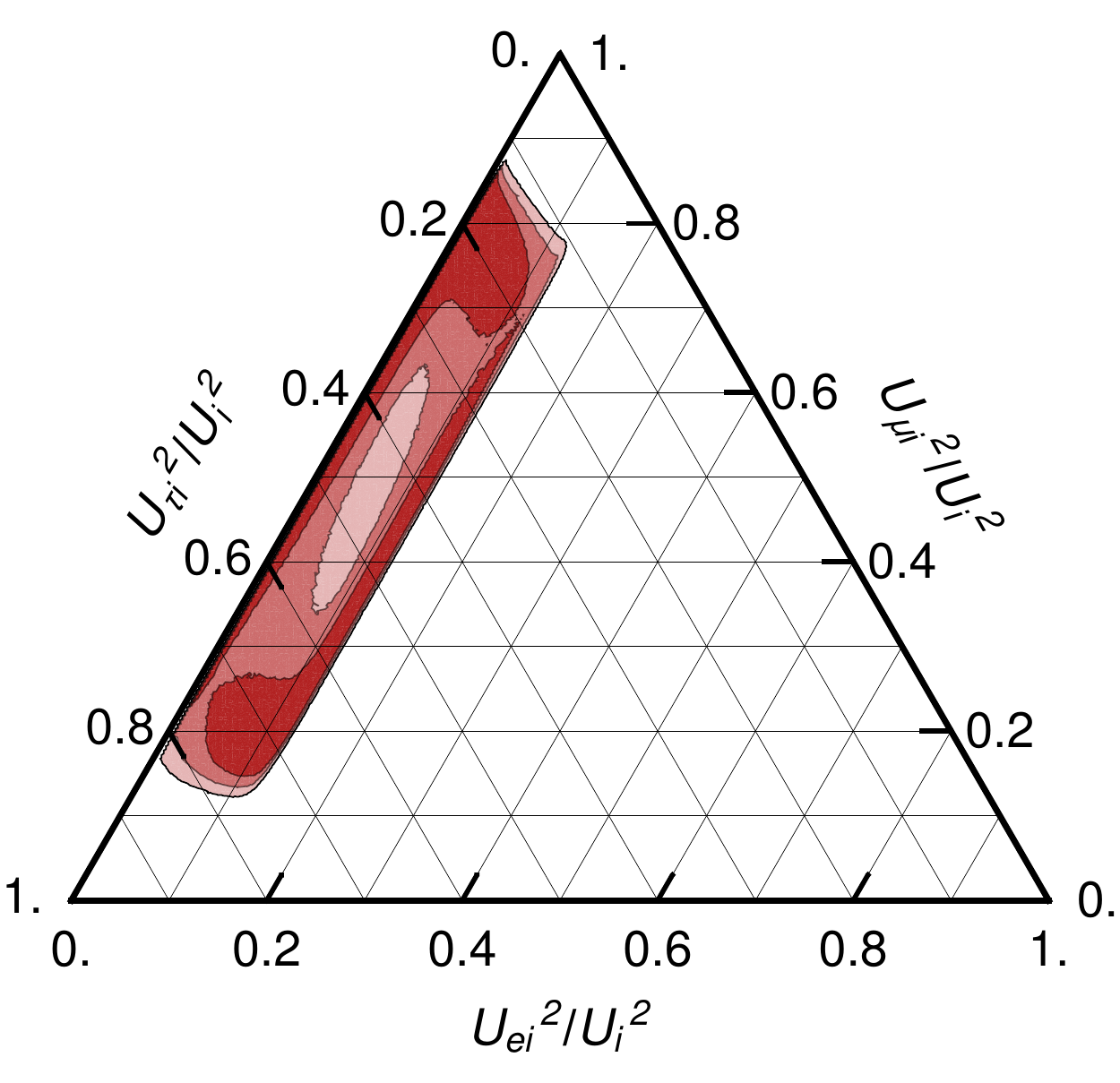}
\caption{Flat prior on $\alpha$, using NuFIT~3.2 data.}
\label{fig:NOalphav32}
\end{subfigure}
\begin{subfigure}{0.5\linewidth}
\includegraphics[width=\linewidth]{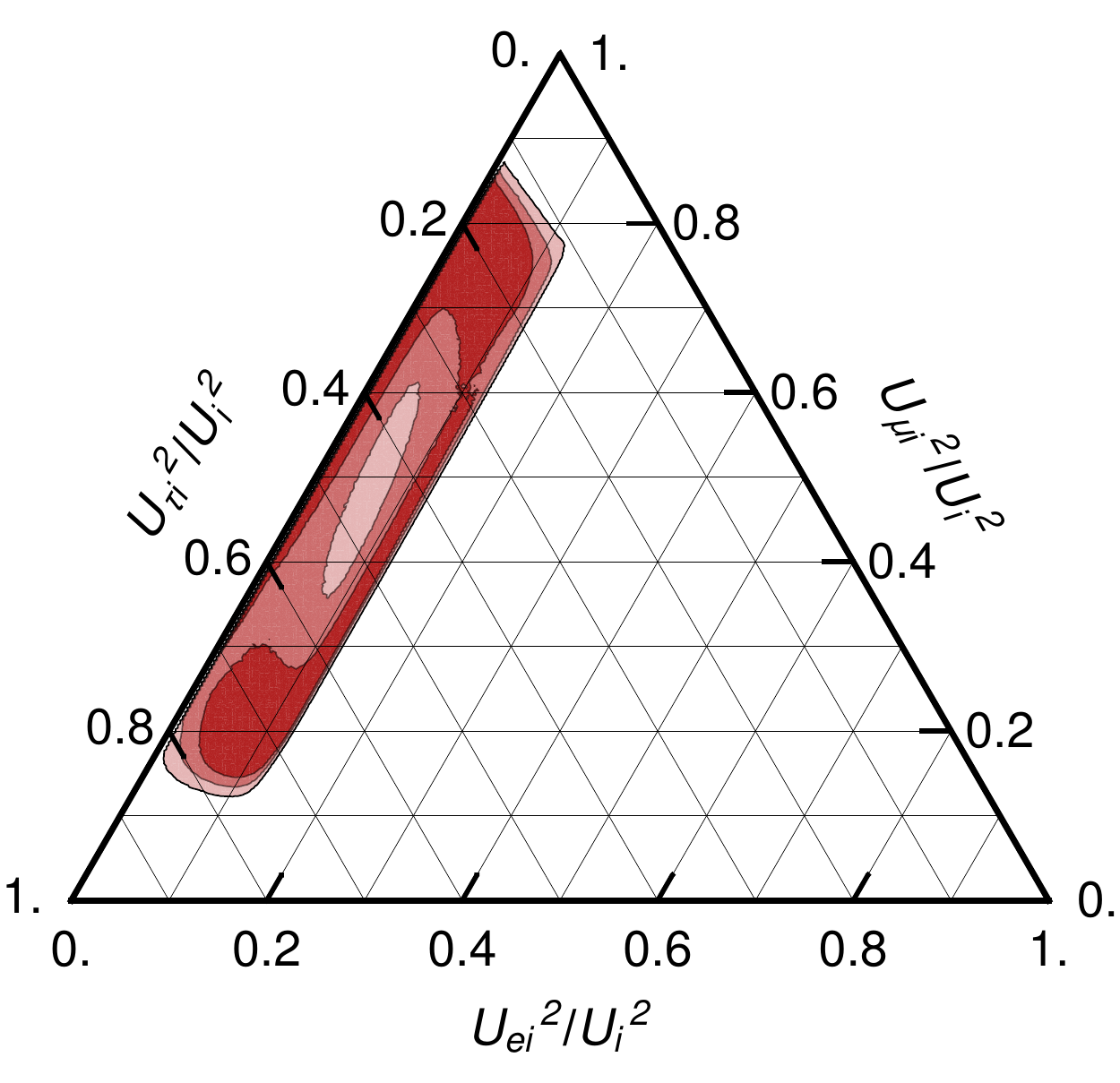}
\caption{Flat prior on $\sin(\alpha/2+\delta)$, using NuFIT~3.2 data.}
\label{fig:NOsinv32}
\end{subfigure}
\caption{The different shades indicate the 1$\sigma$ (darkest), 2$\sigma$ and 3$\sigma$ (lightest) probability contours for the ratios $U_a^2/U^2$ for $n = 2$ and NO that can be obtained from present neutrino oscillation data.
The upper Panels (\subref{fig:NOalpha}, \subref{fig:NOsin}) are based on the results from the NuFIT~3.1 global fit to neutrino oscillation data.
In Panel~(\subref{fig:NOalpha}) we assumed that all values of $\alpha$ are equally valid, in Panel~(\subref{fig:NOsin}) we assumed all values of $\sin(\alpha/2+\delta)$ to be equally likely.
For comparison, the lower two Panels~(\subref{fig:NOalphav32}, \subref{fig:NOsinv32}) are based on the NuFIT~3.2 update, which was published after our analysis was finalised.
\label{fig:Chi2NO}}
\end{figure}

\begin{figure}
\begin{subfigure}{0.5\linewidth}
\includegraphics[width=\linewidth]{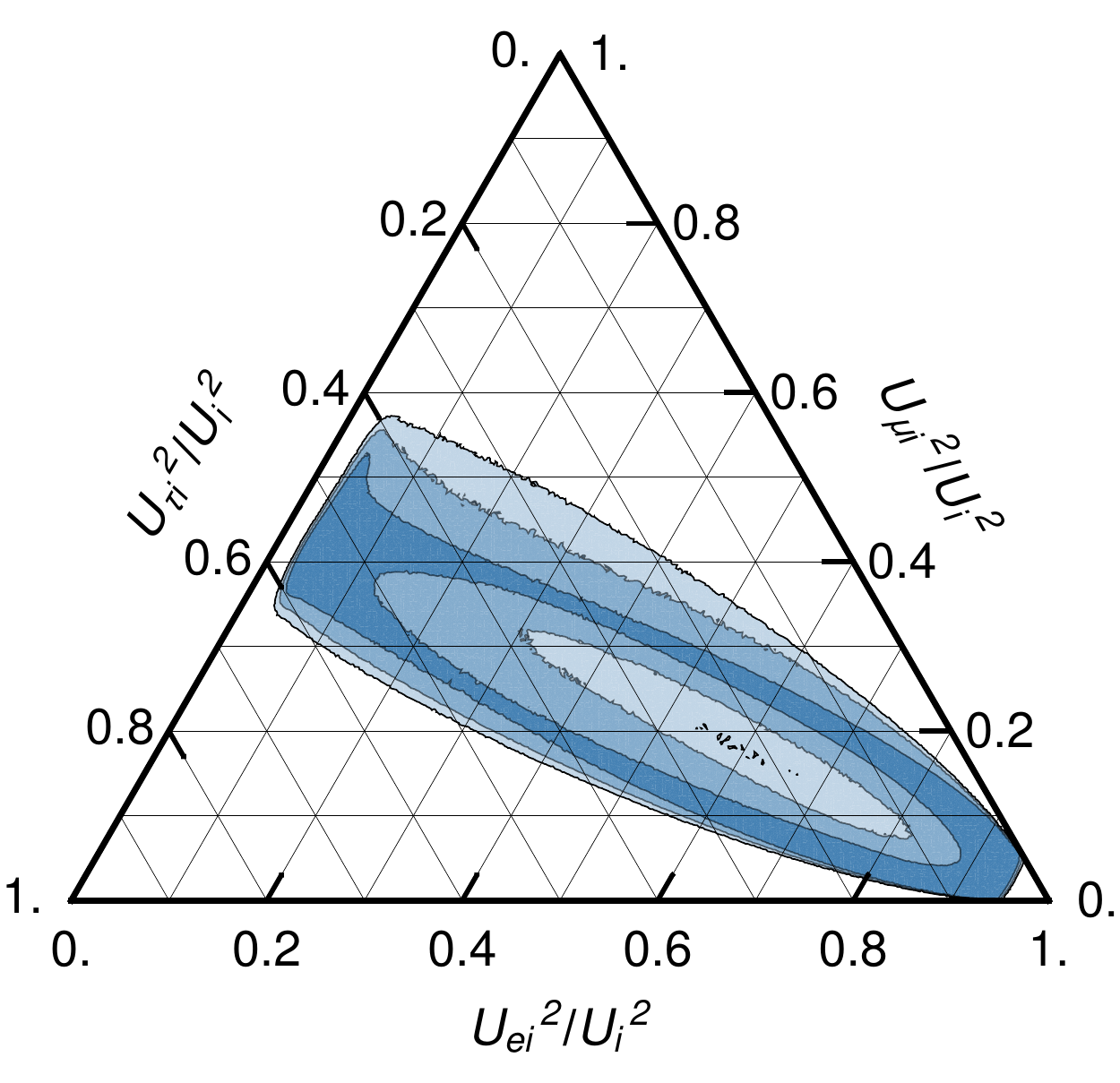}
\caption{Flat prior on $\alpha$.}
\label{fig:normal ordering prior on alpha}
\end{subfigure}
\begin{subfigure}{0.5\linewidth}
\includegraphics[width=\linewidth]{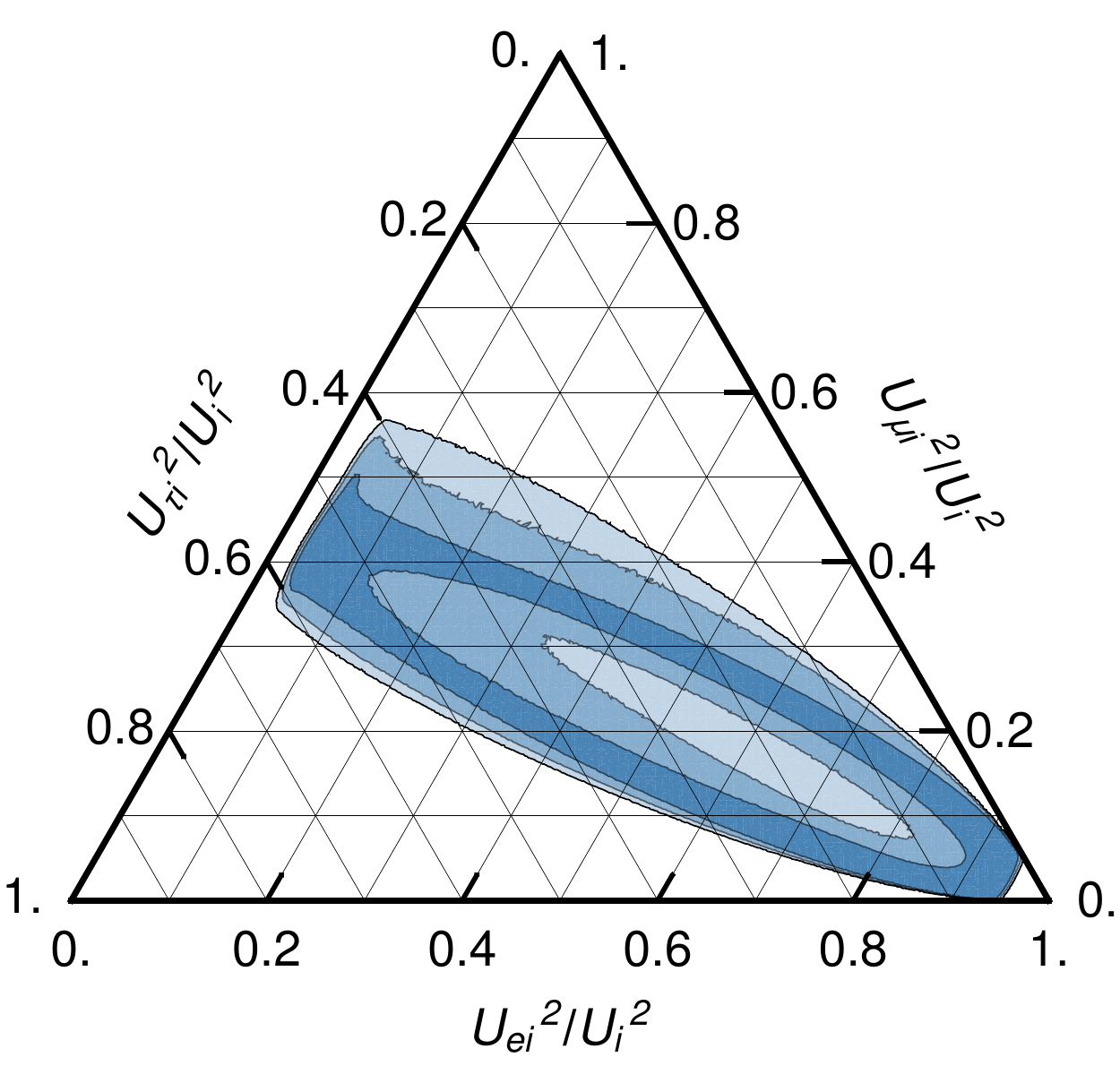}
\caption{Flat prior on $\sin(\alpha/2)$.}
\label{fig:normal ordering prior on sin alpha}
\end{subfigure}
\begin{subfigure}{0.5\linewidth}
\includegraphics[width=\linewidth]{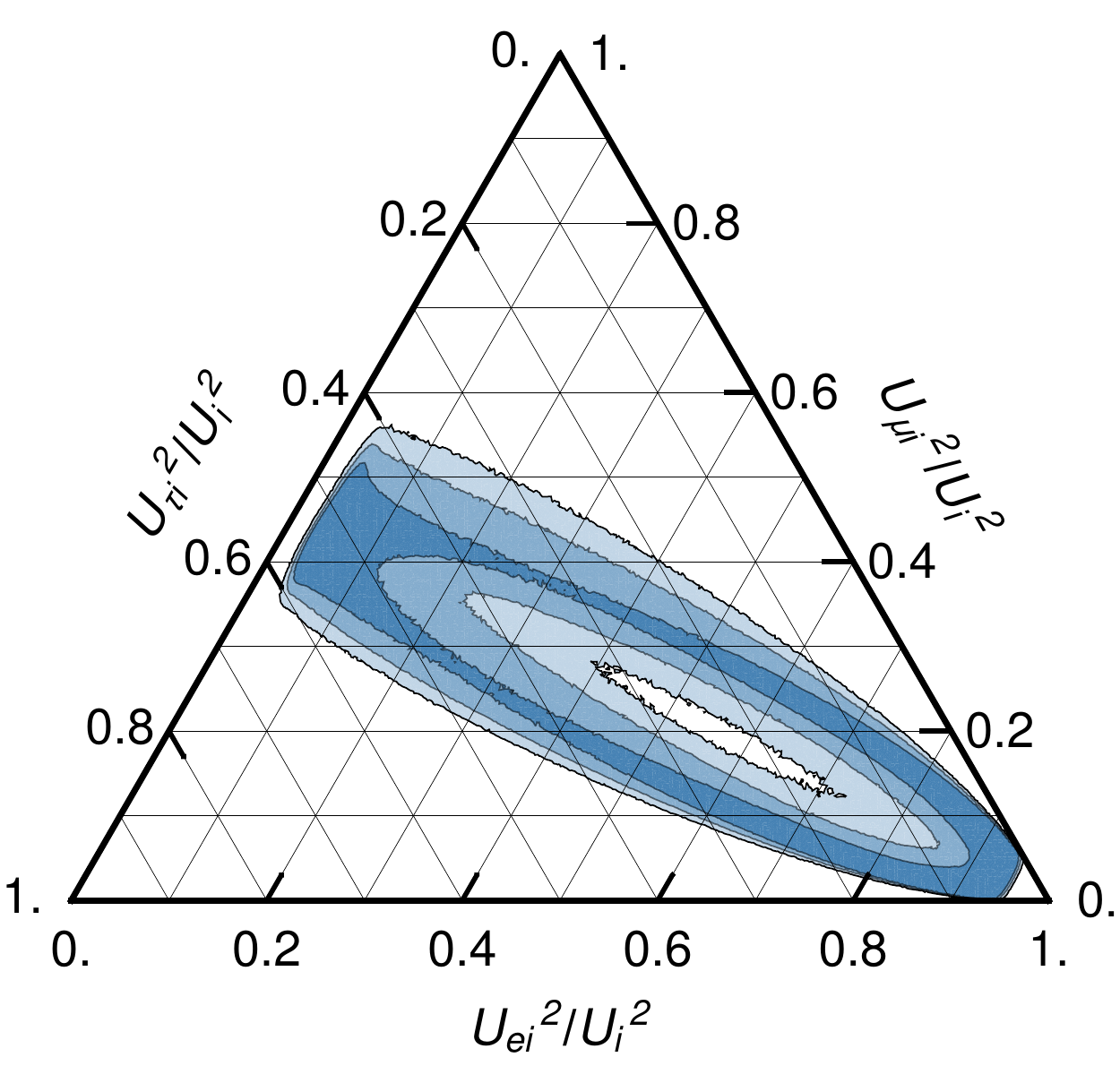}
\caption{Flat prior on $\alpha$, using NuFIT~3.2 data.}
\label{fig:IOalphav32}
\end{subfigure}
\begin{subfigure}{0.5\linewidth}
\includegraphics[width=\linewidth]{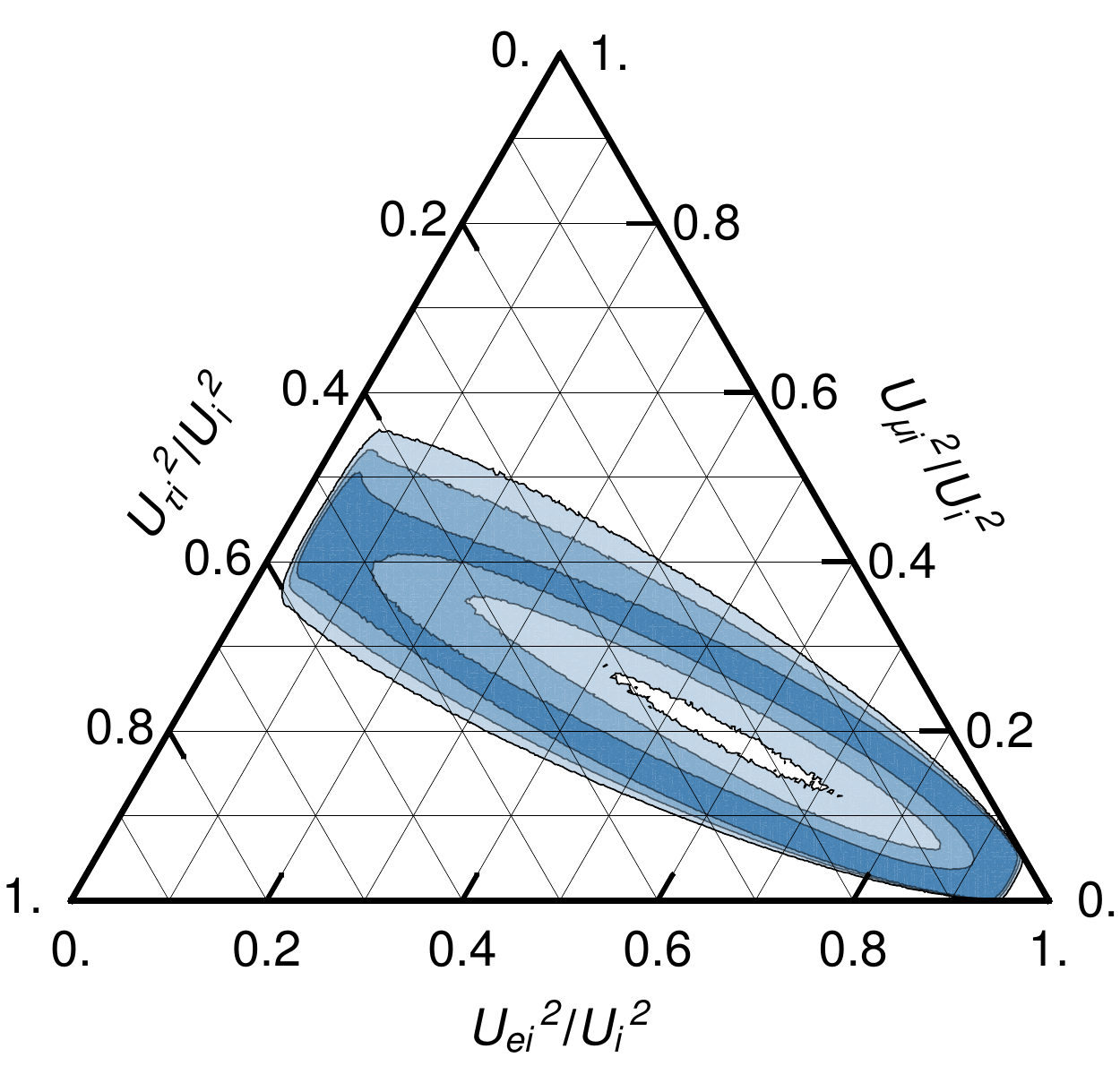}
\caption{Flat prior on $\sin(\alpha/2)$, using NuFIT~3.2 data.}
\label{fig:IOsinv32}
\end{subfigure}
\caption{%
The different shades indicate the 1$\sigma$ (darkest), 2$\sigma$ and 3$\sigma$ (lightest) probability contours for the ratios $U_a^2/U^2$ for $n = 2$ and IO that can be obtained from present neutrino oscillation data.
The upper panels (\subref{fig:normal ordering prior on alpha}) and (\subref{fig:normal ordering prior on sin alpha})  are based on the results from the NuFIT~3.1 global fit to neutrino oscillation data.
In Panel~(\subref{fig:normal ordering prior on alpha}) we assumed that all values of $\alpha$ are equally valid while in Panel~(\subref{fig:normal ordering prior on sin alpha}) we assumed all values of $\sin(\alpha/2)$ to be equally likely.
For comparison, the lower two Panels~(\subref{fig:NOalphav32}, \subref{fig:NOsinv32}) are based on the NuFIT~3.2 update, which was published after our analysis was finalised.}
\label{fig:Chi2IO}
\end{figure}

To identify how favoured each choice of parameters is according to the current neutrino oscillation data we generate a sample of points using a simple implementation of the Metropolis-Hastings algorithm~\cite{Metropolis:1953am, Hastings:1970aa}.
We start with a random choice of the low-energy parameters $x_0=(\uptheta_{12\,0},\:\uptheta_{13\,0},\:\uptheta_{23\,0},\:{\Delta m_{21}^2}_0,\:{\Delta m_{3l}^2}_0,\:{\delta}_0)$.
Each following point in the Markov chain is chosen by generating a candidate point $x^\prime$ from a multivariate normal distribution centred around the last point in the chain $x_i$.
The candidate point is then accepted with the probability
\begin{equation}
    \rho(x_i,x^\prime)
  = \min \left[ \exp(-\frac{\Delta \chi^2(x_i) - \Delta\chi^2(x^\prime)}{2} ) ,\: 1 \right]
\ .
\end{equation}
If the point is accepted, it becomes the next point in the Markov chain, $x_{i+1}=x^\prime$, otherwise we keep $x_{i + 1} = x_i$.
In the limit of large $i$, the density of points reflects the probability distribution in the parameter space in view of the experimental data.
Since $\alpha$ is experimentally unconstrained we have to pick an a priori distribution ``by hand'' from which we chose the samples.
We generate the samples of $\alpha$ in two different ways.
One choice of ``prior'' is a flat distribution in $\alpha$ between 0 and $4\pi$.
In the second approach we use a distribution that is ``flat'' in $U_e^2/U^2$.
This is done by choosing $\alpha$ from a flat distribution in $\sin(\alpha/2+\delta)$ for normal hierarchy and $\sin(\alpha/2)$ for inverted hierarchy according to relations~\eqref{eq:mixing_NOv2} and~\eqref{eq:mixing_IOv2}.
This in principle brings in a dependence on the assumptions (or prior) that one imposes on the value of $\alpha$.
In Figures~\ref{fig:Chi2NO} and~\ref{fig:Chi2IO} we show the likelihoods for the ratios $U_a^2/U^2$ for the two different choices of prior.
The fact that the allowed regions are rather independent of the prior imposed on the unknown parameter $\alpha$ shows that they reflect actual experimental uncertainties (rather than theoretical prejudice on the model).
We compare the results based on the NuFIT~3.1 and NuFIT~3.2 releases.
Though the 3.2 update leads to a visible change in the likelihoods for some of the parameters in $U_\nu$ (in particular $\uptheta_{23}$ and $\delta$), it does not lead to a significant change in the preferred choice of benchmark scenarios.

We finally check whether the above conclusion may  change if constraints from searches for neutrinoless double $\beta$ decay are taken into account.
For $\upmu,\:\upepsilon \ll 1$, the effective Majorana mass $m_{\beta\beta}$ that governs the rate of the decay in the $n = 2$ model can be approximated by~\cite{Asaka:2011pb,Hernandez:2016kel,Drewes:2016lqo}
\begin{subequations}
\begin{align}
    m_{\beta\beta}
  &\simeq
    \abs{\left(1 - f_A(\overline M)\right) m_{\beta\beta}^\nu
        + f_A^2(\overline M) \frac{\overline M^2}{\Lambda^2} \frac{\upmu}{\upepsilon} \abs{\Delta m_\text{atm}} e^{-2 \iu \Re\omega} f}
\ ,
\intertext{with}
 f &= \begin{cases}
      e^{-2 \iu \delta} &\text{for NO}\\
       \cos^2\uptheta_{13} \left(\xi e^{ \iu \alpha_2/2}\sin \uptheta_{12} + \iu e^{ \iu \alpha_1/2}\cos \uptheta_{12}\right)^2 &\text{for IO}
     \end{cases}
\end{align}
\end{subequations}
Here $m_{\beta\beta}^\nu = \sum_i (V_\nu)_{ei}^2m_i \simeq \sum_i (U_\nu)_{ei}^2m_i$ is the contribution from light neutrino exchange.
The function $f_A(M) \simeq \Lambda^2 / \left(\Lambda^2 + \overline M^2 \right)$ contains information about the nuclear matrix elements in the decay.
$\Lambda^2$ is the momentum exchange in the decay, specific values can e.g.~be found in~\cite{Faessler:2014kka}.
For $\upmu = 0$ these expressions reduce to $m_{\beta\beta} = \left(1 - f_A(\overline M)\right) \abs{m_{\beta\beta}^\nu}$, which is always smaller than $\abs{m_{\beta\beta}^\nu}$~\cite{Bezrukov:2005mx}.
Hence, the present non-observation of neutrinoless double $\beta$ decay cannot directly constrain the $U_a^2/U^2$.
For finite $\upmu$ the correction to $m_{\beta\beta}$ is not necessarily small because it scales as $\propto \upmu/\upepsilon$.
This implies that for $\upepsilon \ll 1$, small values of $\upmu$ are favoured by neutrinoless double
$\beta$ decay.
However, since we previously established that the range of allowed $U_{ai}^2/U_i^2$ for $U_i^2$ within reach of NA62 is in good approximation identical to the range that is allowed for $\upmu=0$, present constraints
from neutrinoless double $\beta$ decay do not further reduce the allowed regions in Figure~\ref{fig:AllowedAreas}.

\begin{figure}
\begin{subfigure}{0.5\linewidth}
\includegraphics[width=\linewidth]{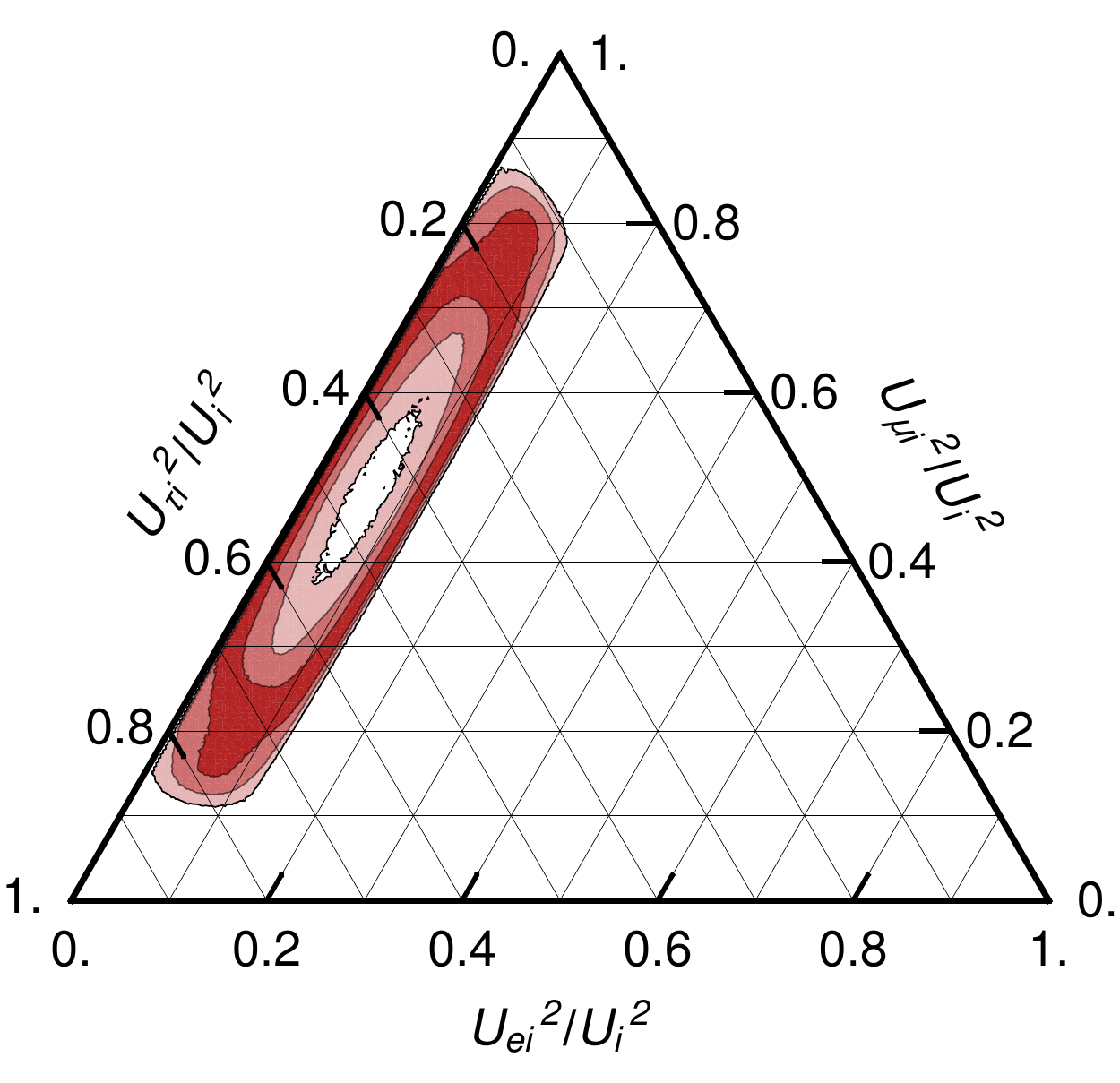}
\caption{Using the NuFIT~3.1 data}
\label{fig:nufit31}
\end{subfigure}
\begin{subfigure}{0.5\linewidth}
\includegraphics[width=\linewidth]{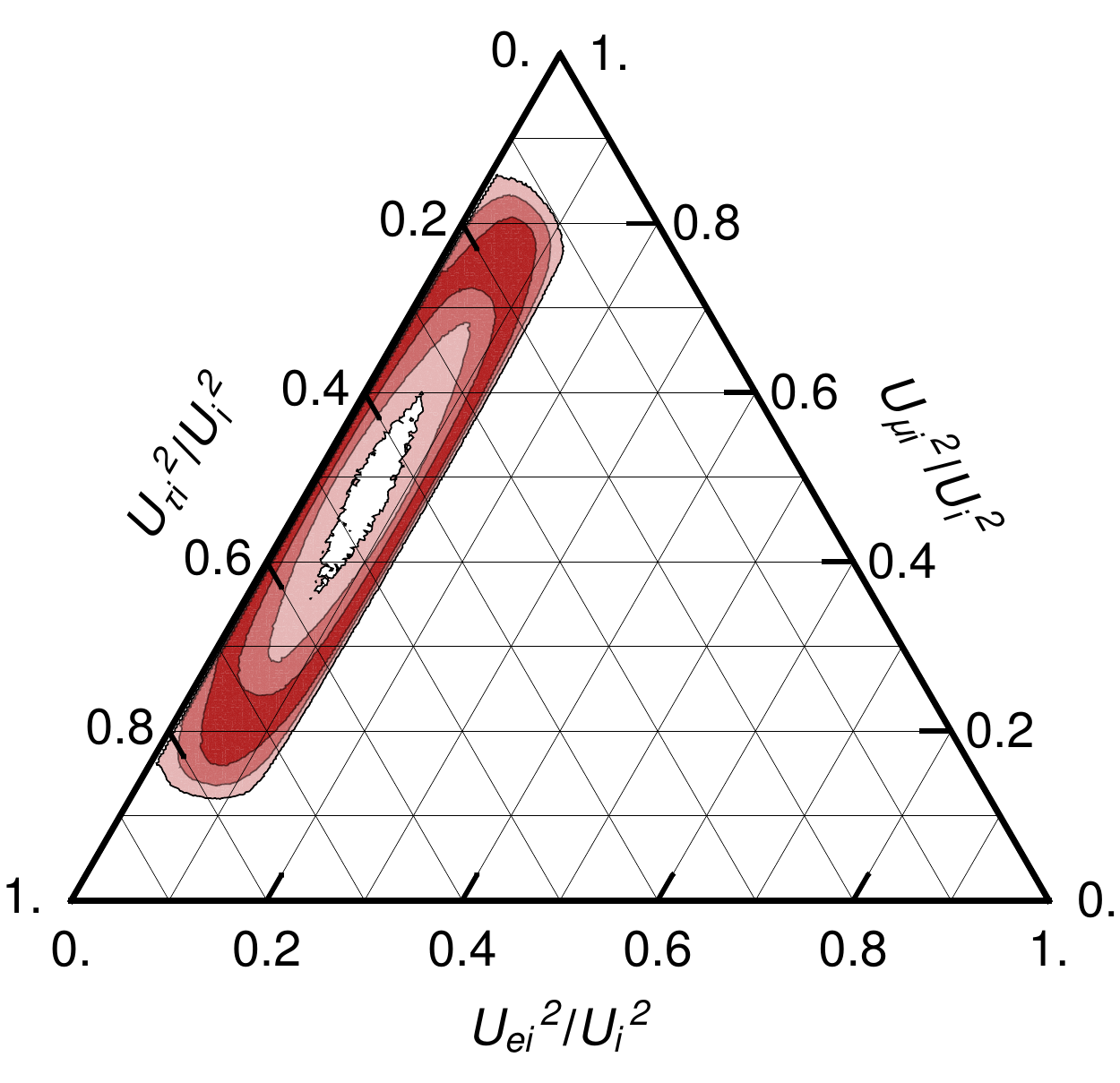}
\caption{Using the NuFIT~3.2 data}
\label{fig:nufit32}
\end{subfigure}
\caption{%
Illustration of the expected improvement in Figure~\ref{fig:Chi2NO} with DUNE~\cite{Acciarri:2015uup}.
Here we assume Gaussian errors around the parameter $\delta=-\pi/2\pm \pi/9$ used as benchmarks in~\cite{DUNE_TAUP_WHITEHEAD_KOERNER}.
For all other parameters we take the one dimensional $\chi^2$ projections from NuFIT~3.1 (\subref{fig:nufit31}) and NuFIT~3.2 (\subref{fig:nufit32}).
We assume that all values of $\alpha$ are equally valid.
In reality we expect even stronger constraints as all low-energy parameters will be measured with a higher precision, we here have only taken into account the expected improvement in $\delta$.}
\label{fig:Chi2NODUNE}
\end{figure}

In summary, the model with $n = 2$ is very predictive as far as the flavour mixing pattern is concerned.
Out of the 11 new parameters in addition to the SM, five are fixed by neutrino oscillation data (cf.~Table~\ref{tab:active_bounds}).
In the symmetric limit, the mass splitting $\Delta M$ vanishes and the phase $\Re \omega$ does not affect the $U_a^2$ at leading order in $\upepsilon$.
By forming the ratios $U_a^2/U^2$, also the dependence on $\overline M$ can be eliminated,
so that these ratios can be determined from fixing the Dirac phase $\delta$ and Majorana phase $\alpha$ in $U_\nu$ alone.
Since $\delta$ is constrained by light neutrino oscillation data, there is only one truly unknown parameter $\alpha$.
Figures~\ref{fig:Chi2NO} and~\ref{fig:Chi2IO} show that the predictions for the allowed range of $U_a^2/U^2$ are rather independent of the theoretical prejudice about the value of $\alpha$.
Hence, they can be used to define benchmark scenarios that mark the corners of the allowed flavour mixing patterns for $n = 2$.
Figure~\ref{fig:Chi2NODUNE} shows how these constraints are expected to improve with the DUNE experiment.%
\footnote{NA62 is expected to probe the range of $U^2$ where the symmetric limit can be applied to determine the allowed values for $U_a^2/U^2$.
It is, however, worthwhile noting that the excellent mass resolution of the experiment allows to resolve the HNL masses for $\upmu < 10^{-2}$~\cite{CortinaGil:2017mqf}.
A measurement of the individual $U_{ai}^2$ would, in combination with a measurement of $\delta$ in neutrino oscillation experiments, in principle allow to extract all parameters in the Lagrangian~\eqref{eq:Lagrangian} for $n = 2$, making this a fully testable model of neutrino masses and baryogenesis~\cite{Drewes:2016jae}.
}

\subsection{The Model with $n = 3$}

The model with $n = 2$ is highly predictive because of its minimality.
For $n = 3$ the flavour mixing patterns are far less restricted.
There is no lower bound on the individual $U_{ai}^2$ from neutrino oscillation data~\cite{Gorbunov:2013dta,Drewes:2015iva}.
It is, for instance, possible to set $F_{e 1} = F_{\mu 1} = 0$ by fixing the entirely unconstrained Majorana phases $\alpha_1$ and $\alpha_2$ for arbitrary choices of all other parameters, including $m_\text{lightest}$.%
\footnote{For $m_\text{lightest} = 0$ it is even possible to set $F_{e 1} = F_{\mu 1} = F_{\tau 1} = 0$, which effectively reduces the model with $n = 3$ to the model with $n = 2$.}
Of course, in this case $N_2$ and $N_3$ have to mix with the first two SM generations in order to explain the mixing of light electron and muon neutrinos.
However, $M_2$ and $M_3$ can be outside the reach of NA62, so that the only observable heavy neutrino can exclusively mix with the third generation, which is impossible for $n = 2$.
This implies that it is not possible to make any reliable predictions for the values of the $U_{ai}^2$ in general.
However, observably large $U_{ai}^2$ still require cancellations in the seesaw relations~\eqref{seesaw}. This is only possible without significant tunings if the $m_i$ are protected by a symmetry.
The implementation of the same generalised $B-L$ symmetry discussed in Section~\ref{sec:nis2} requires that two of the $N_i$ (which we may choose to call $N_2$ and $N_3$) effectively behave like the two heavy neutrinos in the $n = 2$ model, while the third one has much smaller mixings $\sim m_\text{lightest}/M_1$.
This is precisely the behaviour that is observed in the $\nu$MSM, where the feebly coupled neutrino is a DM candidate.
Since only $N_2$ and $N_3$ can realistically be observed at NA62, the considerations from the previous Section~\ref{sec:nis2} also apply to the $n = 3$ case.
The only difference is that, while these are unavoidable predictions in the case $n = 2$, they can be circumvent by "tuning" in the parameters in the scenario with $n = 3$ if one chooses to explain the small $m_i$ by accidental cancellations in relation~\eqref{seesaw}.

\section{Estimate of the NA62 Sensitivity}
\label{sec:sensitivity}

\subsection{The NA62 Experiment}
\label{sec:NA62}

The main goal of the NA62 experiment~\cite{NA62:2017rwk} which is currently taking data at the CERN SPS is to measure the Branching Ratio (BR) of the $K^+ \to \pi^+ \nu \overline \nu$ decay with a precision of at least \unit[10]{\%}.
In order to achieve this goal the experiment needs to collect about $10^{13}$ kaon decays of which $\order{\text{few} \times 10^{12}}$ have already been collected in the current run~\cite{pbc}.

In its normal operation mode, the \emph{kaon mode}, the primary \unit[400]{GeV} proton beam impinges
on a \unit[400]{mm} long cylindrical beryllium target with a diameter of \unit[2]{mm} which is used to produce a secondary positively charged hadron beam with a momentum of \unit[75]{GeV}.
\unit[100]{m} downstream of the target the secondary beam reaches the \unit[120]{m} long evacuated decay volume which has a diameter of \unit[2]{m}.
About \unit[6]{\%} of the hadron beam are kaons, which are identified and timestamped by a N\textsubscript{2} filled Cherenkov counter located along the beam line.
Three silicon pixel stations measure momentum and time of all the particles in the beam at a rate
of \unit[750]{MHz}.
A guard ring detector tags hadronic interactions in the last pixel station at the entrance of the decay volume.
Large angle electromagnetic calorimeters made of lead glass blocks surrounding the decay vessel are used to veto particles up to \unit[50]{mrad}.
A magnetic spectrometer made of straw tubes in vacuum measures the momentum of the charged particles.
A Ring-Imaging Cherenkov (RICH) counter filled with Neon separates $\pi$, $\mu$ and $e$ for momenta up to \unit[40]{GeV}.
The time of flight for charged particles is measured both by the RICH and by the scintillator hodoscopes placed downstream of the RICH.
An electromagnetic calorimeter covers the forward region and complements the RICH for the particle identification.
The hadronic calorimeter provides further separation between $\pi$ and $\mu$ based on hadronic energy and a fast scintillator array identifies muons with sub-nanosecond time resolution.

At the nominal beam intensity of $3\times 10^{12}$ protons per pulse, with pulses of \unit[4.8]{s}, the NA62 experiment can collect up to $3\times 10^{18}$ protons on target (POT) per year.

When the experiment is operated in the \emph{dump mode}, the target is pulled up and the primary proton beam
is send directly onto the Cu-Fe based collimators that act as a hadron stopper (or \emph{dump}) located \unit[20]{m} downstream of the target.
In this configuration, about $2 \times 10^{15}$ $D$-mesons and $\sim 10^{11}$ $b$-hadrons are produced from the $10^{18}$ POT, which correspond to about 80 days of data taking at the nominal NA62 beam intensity.
This dataset will be collected during Run 3 (2021--2023) and is assumed to produce the sensitivity plots discussed in Section~\ref{sec:sensitivity}.

\subsection{Benchmark scenarios}

\begin{table}
\begin{subtable}{0.5\linewidth}
\centering
\begin{tabular}{cc@{ : }c@{ : }cccc}
 \toprule
 & \multicolumn{3}{c}{Ratio}
 & \multicolumn{3}{c}{Percent}
 \\ \cmidrule(r){2-4}
 \cmidrule(l){5-7}
 & $U_{e i}^2$
 & $U_{\mu i}^2$
 & $U_{\tau i}^2$
 & $U_{e i}^2$
 & $U_{\mu i}^2$
 & $U_{\tau i}^2$
 \\ \midrule
 A)
 & 1
 & 160
 & 27.8
 & 0.530
 & 84.7
 & 14.7
 \\ B)
 & 1
 & 1.71
 & 5.62
 & 12.0
 & 20.5
 & 67.5
 \\ C)
 & 1
 & 10.5
 & 15.9
 & 3.65
 & 38.3
 & 58.0
 \\ D)
 & 1
 & 0
 & 0
 & 100
 & 0
 & 0
 \\ E)
 & 0
 & 1
 & 0
 & 0
 & 100
 & 0
 \\ F)
 & 0
 & 0
 & 1
 & 0
 & 0
 & 100
 \\ \bottomrule
\end{tabular}
\caption{Benchmark scenarios used in this analysis.}
\label{tab:our benchmark scenarios}
\end{subtable}
\begin{subtable}{0.5\linewidth}
\centering
\begin{tabular}{cc@{ : }c@{ : }cccc}
 \toprule
 & \multicolumn{3}{c}{Ratio}
 & \multicolumn{3}{c}{Percent}
 \\ \cmidrule(r){2-4}
 \cmidrule(l){5-7}
 & $U_{e i}^2$
 & $U_{\mu i}^2$
 & $U_{\tau i}^2$
 & $U_{e i}^2$
 & $U_{\mu i}^2$
 & $U_{\tau i}^2$
 \\ \midrule
 1)
 & 52
 & 1
 & 1
 & 96.3
 & 1.85
 & 1.85
 \\ 2)
 & 1
 & 16
 & 3.8
 & 4.8
 & 76.9
 & 18.3
 \\ 3)
 & 0.061
 & 1
 & 4.3
 & 1.14
 & 18.7
 & 80.2
 \\ 4)
 & 48
 & 1
 & 1
 & 96
 & 2
 & 2
 \\ 5)
 & 1
 & 11
 & 11
 & 4.35
 & 47.83
 & 47.83
 \\ \bottomrule
 \\
\end{tabular}
\caption{Benchmark scenarios used in the SHiP analysis.}
\label{tab:ship benchmark scenarios}
\end{subtable}
\caption{We consider six benchmark scenarios~A)--F) listed in Panel~(\subref{tab:our benchmark scenarios}).
For comparison we also list the benchmark scenarios used in~\cite{Graverini:2214085} for the analysis for the SHiP experiment in Panel~(\subref{tab:ship benchmark scenarios}).
\label{tab:benchmark scenarios}
}
\end{table}

As for any fixed target experiment, the sensitivity of NA62 can only be computed for fixed ratios $U_{e i}^2 : U_{\mu i}^2 : U_{\tau i}^2$. In the symmetric limit this can be used exchangeably with $U_e^2 : U_\mu^2 : U_\tau^2$, cf.~relation~\eqref{SymmLimitCouplings}.
We consider the benchmark scenarios listed in Table~\ref{tab:our benchmark scenarios}.
The scenarios~A)--D) extremise ratios of $U_a^2$ in the symmetric limit.
Scenario~A) minimises the mixing with the first generation and maximises it with the second generation. In the $n = 2$ model it can be realised for $(\alpha,\:\delta,\:\xi) = (-\pi,\:\pi,\:1)$ with NO.
Scenario~B) maximises the mixing with the first generation and minimises it with the second generation for NO.
In the $n = 2$ model it can be realised for $(\alpha,\:\delta,\:\xi) = (\pi,\:\pi,\:1)$.
Scenario~C) minimises the mixing with the first generation for IO.
In the $n = 2$ model it can be realised for $(\alpha,\:\delta,\:\xi) = (-\pi,\:\pi,\:1)$.
The scenarios are marked by stars in Figure~\ref{fig:AllowedAreas}.
We have not included the other "corners" of the allowed regions because they either require $\delta = 0$ for $n = 2$, which is disfavoured by current neutrino oscillation data, or are phenomenologically very similar to one of our scenarios.
We practically use scenario~D) to model the maximal mixing with the first generation for IO.

The scenarios~A)--D) may be compared to the choices 1)--5) listed in Table~\ref{tab:ship benchmark scenarios} and used in the estimates for the SHiP experiment~\cite{Anelli:2015pba,Alekhin:2015byh} in reference~\cite{Graverini:2214085}, which were introduced in references~\cite{Gorbunov:2007ak,Canetti:2010aw}.
Scenarios~$1)$--$3)$ maximise $U_e^2$, $U_\mu^2$ and $U_\tau^2$ for NO in view of the constrains from neutrino oscillation experiments at the time when they were proposed.
However, such large $U_\tau^2$ are disfavoured by more recent data, while a stronger contributions from $U_e^2$ and $U_\mu^2$ now seem to be allowed.
Scenario~$4)$, which at the time was motivated by IO, is very similar to scenario~$1)$, while scenario~$5)$ is similar to our scenario~C).
Hence, the scenarios~A)--D) can be seen as updates to the scenarios~$1)$--$5)$ in view of the most recent neutrino oscillation data (in particular the measurement of $\uptheta_{13}$, which was unknown at the time when the articles in references~\cite{Gorbunov:2007ak,Canetti:2010aw} were written).

The scenarios~D)--F) are the extreme cases in which a $N_i$ couples only to one SM generation.
Scenarios~E) and F) are not allowed in the $n = 2$ model for $U_{ai}^2$ within reach of NA62, and in the $n = 3$ model they can only be realised with significant tuning.
It is nevertheless instructive to include them in order to understand the most optimistic and most pessimistic predictions for the NA62 sensitivity without theoretical prejudice.
As mentioned before, scenario~D) can almost be realised for IO and may therefore be classified as "allowed".

\subsection{NA62 sensitivity computation}

The computation of the NA62 sensitivity for $U^2_{ia}$ ($a=e,\:\mu,\:\tau$) in different scenarios has been performed using a toy Monte Carlo in which all the kinematics of the $N_i$ production and decay processes have been implemented and the geometrical acceptance for the decay products has been properly evaluated using the geometry of the experiment as described in~\cite{NA62:2017rwk}.

Heavy neutrinos can be produced in hadron decays where a SM neutrino $\nu_a$ is replaced by a $N_i$ through the mixing $\theta_{ai}$.
At the SPS energies this happens via production of strange, charm and beauty hadrons.
Above the kaon mass, the main production processes are the decays of charm and beauty hadrons which are kinematically allowed.
The number $n_N$ of $N_i$  produced in a beam-dump experiment can be quantified by
\begin{equation}
    n_N
  \simeq
    2 N_\text{POT}
    \left(
        \chi_c \sum_{D_j = D^+,\:\mathrlap{D^0,\:D_s}} f_{D_j} \BR \left(D_j \to X N_i\right)
      + \chi_b \sum_{B_k = B^+,\:\mathrlap{B^0,\:B_s}} f_{B_k} \BR \left(B_K \to X N_i \right)
    \right)
\ ,
\label{eq:N}
\end{equation}
where $N_\text{POT}$ is the total number of protons on target, $\chi_c$ and  $\chi_b$ are the ratios of the production cross sections of $c$ and $b$ quarks with respect to the total $pp$ cross section for a thick target%
\footnote{The factor 2 takes into account the fact that charm and beauty quarks are always produced in pairs.},
$f_{D_j}$ and $f_{B_k}$ are the production fractions of charm or beauty mesons and finally $\BR\left(D_{j} \to  X N_i\right)$ and $\BR\left(B_{k} \to X N_i\right)$ are the branching fractions of charm or beauty mesons into heavy
neutrinos%
\footnote{The inclusion of charge-conjugated processes is implied throughout this document.},
which depend on their mass $M_i$ and on the coupling parameter $U^2_i$.

In the sensitivity computation we assume $10^{18}$ POT, as discussed in Section~\ref{sec:NA62}.
The $c$- and $b$-hadrons can be created in the hadronisation process originated from primary protons, and from all secondary products of the hadronic shower in the dump, such as protons, neutrons, and pions.
Given the ratio $\chi_c / \chi_b \sim 10^4$, the $N_i$ production via charm decays is the dominant process up to the $D$-meson masses, while the contribution from $b$-hadrons decays at the NA62 intensity is almost negligible.
The composition of the shower and the kinematics of the produced $c$- and $b$-hadrons have been studied by simulating the \unit[400]{GeV} proton beam on a thick ($\sim 11$ interaction lengths) high-$Z$ target with \texttt{Pythia}~6.4~\cite{Sjostrand:2006za}.

\begin{savenotes}
\begin{figure}
\centering
\includegraphics[width=.8\linewidth]{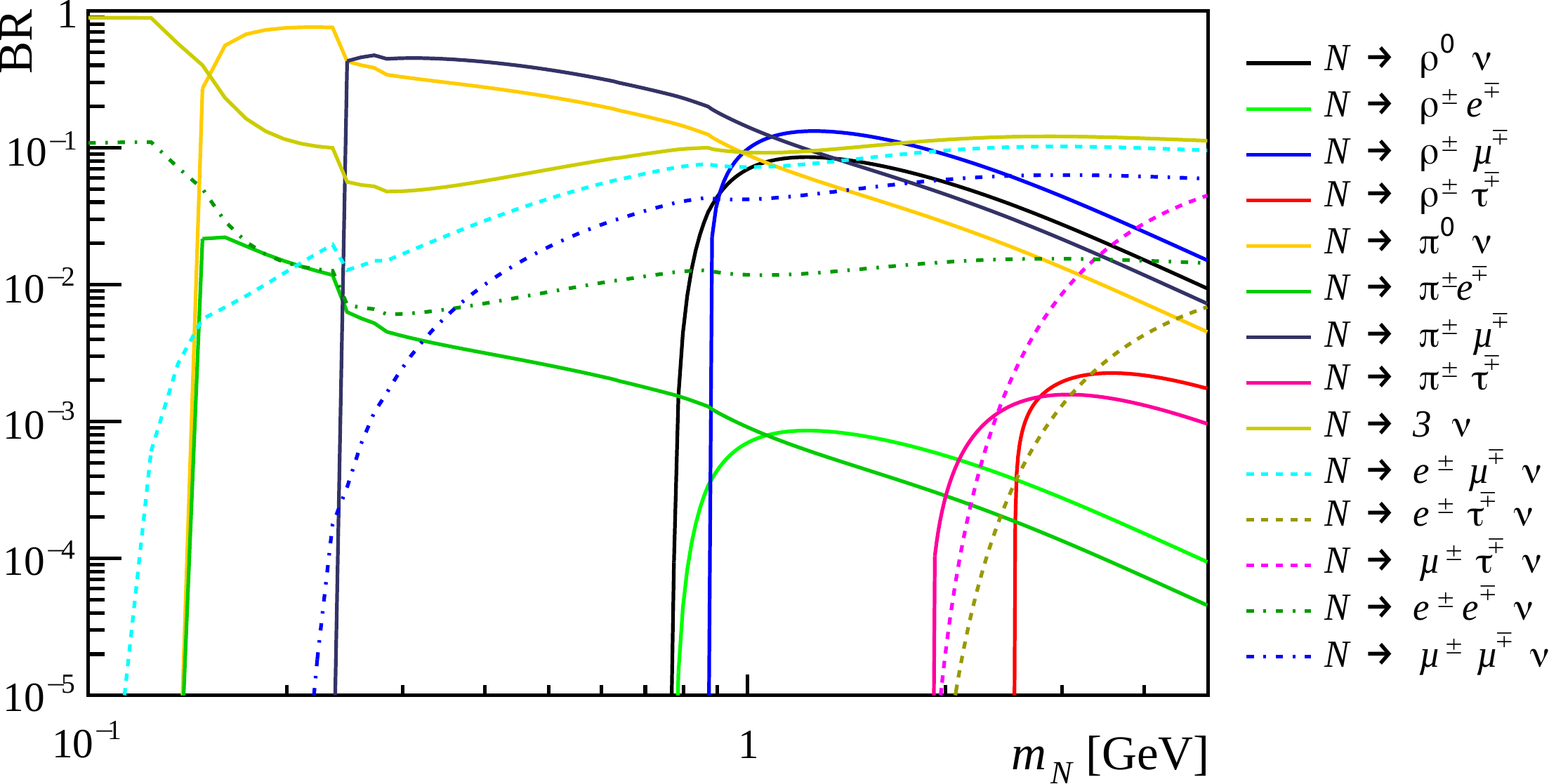}
\caption[]{%
HNL branching fractions used in this work as a function of the HNL mass for scenario~A) ($U^2_{ie} : U^2_{i\mu} : U^2_{i\tau} = 1:160:27.8$) as given in~\cite{Gorbunov:2007ak}.%
\footnote{While this paper was under review, an updated computation of these rates had appeared in~\cite{Bondarenko:2018ptm}.}
}
\label{fig:BRdecay_160}
\end{figure}
\end{savenotes}

Once produced, the $N_i$ decay to SM particles via their $\theta$-suppressed weak interactions.
These massive neutrino states can decay to a variety of final states through charged- and neutral-current processes.
The main decay channels of $N_i$ with masses below a few GeV are
\begin{equation}
    N_i
  \to
3 \nu ,\:
\pi^0 \nu ,\:
\pi^{\pm} \ell^{\mp} ,\:
\rho^0 \nu ,\:
\rho^{\pm} l ,\:
\ell^+ \ell^- \nu
\ , \label{eqn:decays}
\end{equation}
where $\ell = e,\:\mu,\:\tau$.
The branching fractions as a function of the $N_i$ mass are shown for scenario~A) in Figure~\ref{fig:BRdecay_160}.
The relative weight of the branching fractions depends on the scenario under consideration.
The analytical formulae can be found in~\cite{Gorbunov:2007ak}.

\begin{figure}
\begin{subfigure}{.49\linewidth}
\tikzsetnextfilename{NA62_results}\input{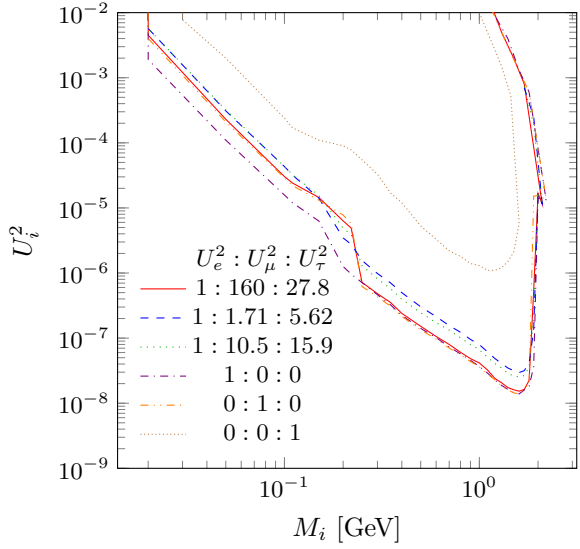}
\begin{tikzpicture}
\begin{loglogaxis}[
ylabel = {$U^2_i$},
enlarge x limits = 0.077,
legend pos = south west,
cycle list = {
{red,solid},
{blue,dashed},
{green!70!black,dotted},
{violet, dashdotted},
{orange, dashdotdotted},
{brown, densely dotted}
},
]
\addlegendimage{empty legend}\addlegendentry{$U^2_e : U^2_\mu : U^2_\tau$}
\addplot file {data/NA62_sc160.csv};\addlegendentry{$1 : 160 : 27.8$}
\addplot file {data/NA62_sc1_71.csv};\addlegendentry{$1 : 1.71 : 5.62$}
\addplot file {data/NA62_sc10_5.csv};\addlegendentry{$1 : 10.5 : 15.9$}
\addplot file {data/Ue-NA62.csv};\addlegendentry{$1 : 0 : 0$}
\addplot file {data/Umu-NA62.csv};\addlegendentry{$0 : 1 : 0$}
\addplot file {data/Utau-NA62.csv};\addlegendentry{$0 : 0 : 1$}
\end{loglogaxis}%
\end{tikzpicture}%
\caption{Scenarios~A)--F) in NA62.}
\label{fig:result all scenarios}
\end{subfigure}
\hfill
\begin{subfigure}{.49\linewidth}
\tikzsetnextfilename{U2e}\input{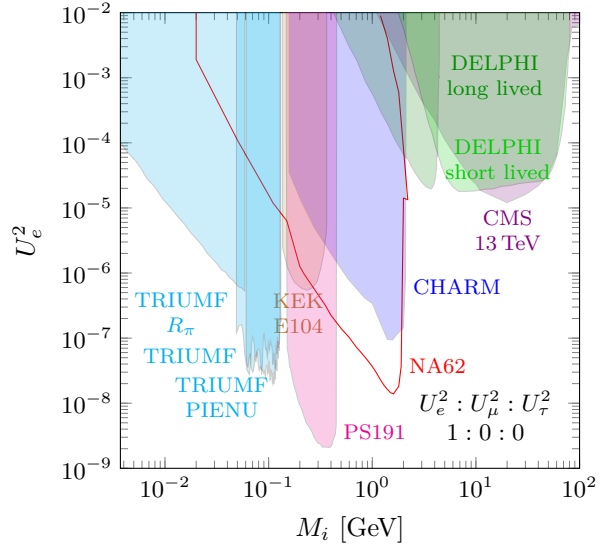}
\newcommand{\past}[5]{\pastplot{Ue-#1}{#2}{#3}{#4}{#5}}
\newcommand{\present}[5]{\presentplot{Ue-#1}{#2}{#3}{#4}{#5}}
\begin{tikzpicture}
\begin{loglogaxis}[ylabel={$U^2_e$}]
\addlegendimage{empty legend}\addlegendentry{$U^2_e : U^2_\mu : U^2_\tau$}
\addlegendimage{empty legend}\addlegendentry{$1:0:0$}
\present{NA62}{NA62}{red}{0.55}{right}

\past{CHARM}{CHARM}{blue}{0.4}{right}
\past{CMS-13TeV}{CMS\\\unit[13]{TeV}}{violet}{0.37}{below}
\past{DELPHI-longlived}{DELPHI\\long lived}{green!50!black}{0.95}{right}
\past{DELPHI-shortlived}{DELPHI\\ short lived}{green!80!black}{0.55}{above}
\past{KEK}{KEK\\E104}{brown}{0.35}{below}
\past{TRIUMF-low}{TRIUMF\\$R_\pi$}{cyan}{0.95}{below left}
\past{TRIUMF-high}{TRIUMF}{cyan}{0.1}{left}
\past{TRIUMF-PIENU}{TRIUMF\\PIENU}{cyan}{0.7}{below left}
\past{PS191}{PS191}{magenta}{0.53}{right}
\end{loglogaxis}%
\end{tikzpicture}%
\caption{Comparison for pure $U_e^2$ coupling.}
\label{fig:result e}
\end{subfigure}
\\[2ex]%
\begin{subfigure}{.49\linewidth}
\tikzsetnextfilename{U2mu}\input{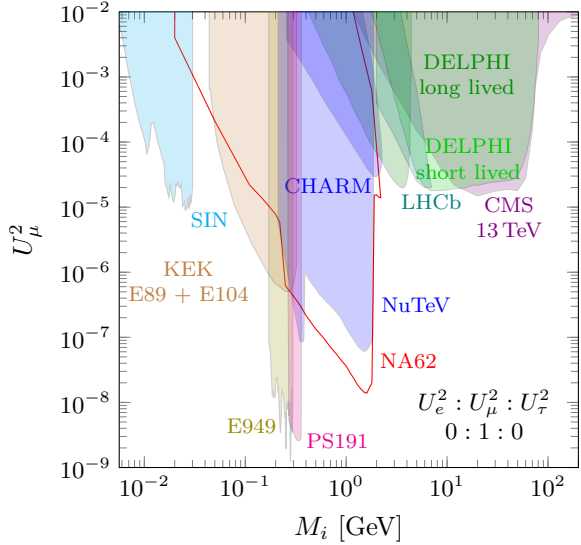}
\newcommand{\past}[5]{\pastplot{Umu-#1}{#2}{#3}{#4}{#5}}
\newcommand{\present}[5]{\presentplot{Umu-#1}{#2}{#3}{#4}{#5}}
\begin{tikzpicture}
\begin{loglogaxis}[ylabel={$U^2_\mu$}]
\addlegendimage{empty legend}\addlegendentry{$U^2_e : U^2_\mu : U^2_\tau$}
\addlegendimage{empty legend}\addlegendentry{$0:1:0$}
\present{NA62}{NA62}{red}{0.53}{right}

\past{CHARM}{CHARM}{blue}{0.52}{below left}
\past{CMS-13TeV}{CMS\\\unit[13]{TeV}}{violet}{0.42}{below}
\past{DELPHI-longlived}{DELPHI\\long lived}{green!50!black}{0.95}{right}
\past{DELPHI-shortlived}{DELPHI\\ short lived}{green!80!black}{0.55}{above}
\past{E949}{E949}{olive}{0.37}{below left}
\past{KEK}{KEK\\ E89 + E104}{brown}{0.7}{below left}
\past{SIN}{SIN}{cyan}{0.84}{below right}
\past{LHCb-DisVert}{LHCb}{teal}{0.25}{below}
\past{NuTeV}{NuTeV}{blue}{0.8}{right}
\past{PS191}{PS191}{magenta}{0.5}{right}
\end{loglogaxis}%
\end{tikzpicture}%
\caption{Comparison for pure $U_\mu^2$ coupling.}
\label{fig:result mu}
\end{subfigure}
\hfill
\begin{subfigure}{.49\linewidth}
\tikzsetnextfilename{U2tau}\input{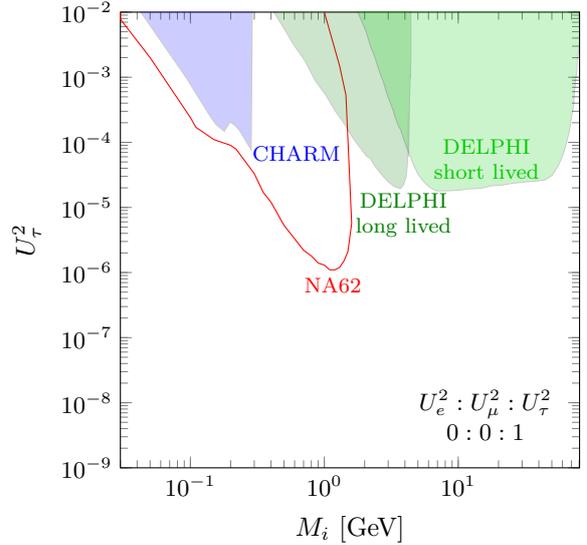}
\newcommand{\past}[5]{\pastplot{Utau-#1}{#2}{#3}{#4}{#5}}
\newcommand{\present}[5]{\presentplot{Utau-#1}{#2}{#3}{#4}{#5}}

\begin{tikzpicture}
\begin{loglogaxis}[ylabel={$U^2_\tau$}]
\addlegendimage{empty legend}\addlegendentry{$U^2_e : U^2_\mu : U^2_\tau$}
\addlegendimage{empty legend}\addlegendentry{$0:0:1$}
\present{NA62}{NA62}{red}{0.52}{below}

\past{CHARM}{CHARM}{blue}{0.535}{below right}
\past{DELPHI-longlived}{DELPHI\\long lived}{green!50!black}{0.475}{below}
\past{DELPHI-shortlived}{DELPHI\\ short lived}{green!80!black}{0.55}{above}
\end{loglogaxis}%
\end{tikzpicture}%
\caption{Comparison for pure $U_\tau^2$ coupling.}
\label{fig:result tau}
\end{subfigure}
\caption{%
Sensitivity of the NA62 experiment in the scenarios~A)--F) in Panel~(\subref{fig:result all scenarios}).
The region above the curves marks the expected exclusion regions for 2.3 events for each scenario.
For the three extreme cases D), E) and F) we compare our results with current exclusion limits 
in the Panels~(\subref{fig:result e}), (\subref{fig:result mu}) and (\subref{fig:result tau}), respectively.
The strongest bounds come from SIN~\cite{Abela:1981nf}, TRIUMF~\cite{Britton:1992pg, Britton:1992xv, PIENU:2011aa}, KEK~\cite{Yamazaki:1984sj}, E949~\cite{Artamonov:2014urb}, PS191~\cite{Bernardi:1987ek}, CHARM~\cite{Bergsma:1985is}, NuTeV~\cite{Vaitaitis:1999wq, Vaitaitis:2000vc}, DELPHI~\cite{Abreu:1996pa}, LHCb~\cite{Aaij:2016xmb, Antusch:2017hhu} and CMS~\cite{Sirunyan:2018mtv}.
}
\label{fig:result}
\end{figure}

The NA62 detector is able to reconstruct all final states with two charged tracks.
The number of events reconstructed in the NA62 detector is given by
\begin{equation}
    N_\text{obs}
  = \sum_{\mathllap{I=\text{prod}}\text{uction m}\mathrlap{\text{odes}}} n_{N,I}
    \sum_{f,\:f'\mathrlap{=e,\:\mu,\:\tau,\:\pi,\:K}}
    \BR \left(N_i \to f^+ f^{\prime-} X \right)
    \mathcal A_i\left( f^+ f^{\prime-} X,\: M_i, U^2_{e,\:\mu,\:\tau}\right)
    \varepsilon \left(f^+ f^{\prime-} X,\: M_i \right)
\ ,
\label{eq:Nobs}
\end{equation}
where $n_{N,I}$ is the number of produced $N_i$ for a given production process $I$, via decays of charm and beauty hadrons as shown in~\eqref{eq:N}, $\mathcal A_I (f^+ f^{\prime-} X,\: m_N,\: U^2_{e,\:\mu,\:\tau})$ is the geometrical acceptance for a $N_i$ produced in the process $I$ of a given mass $M_i$ and coupling $U^2_{e,\:\mu,\:\tau}$ which decays into a final state with two charged tracks ($f^+ f^{\prime-}$) and other decay products $X$ as, for example, photons and neutrinos.
The efficiency $\varepsilon (f^+ f^{\prime-} X)$ is the product of the trigger, reconstruction and selection efficiencies for a given final state, and is currently assumed to be \unit[100]{\%}.
The curves shown in Figure~\ref{fig:result} can therefore slightly change when all the experimental effects are taken into account.%
\footnote{In the conservative case of an overall efficiency of \unit[50]{\%}, the curves would move upwards by a factor $\sqrt 2$.}

The position of the decay vertex of the $N_i$ is required to be inside the \unit[75]{m} long fiducial volume, located \unit[80]{m} downstream of the dump, and within a circle with \unit[1]{m} radius in the transverse plane.
The decay products are required to be within acceptance of the charged hodoscope.
The geometrical acceptance is given by the convolution of the probability that the $N_i$ decays inside the fiducial volume with the probability  that the two charged tracks in the final state are reconstructed in the magnetic spectrometer.

Monte Carlo techniques are used to assess the sensitivity contours in the mass-coupling parameter space.
This is done by generating heavy neutrinos  with different mass and coupling values, letting them decay in the kinematically accessible final states, and computing the corresponding geometrical acceptance.

The exclusion limits at \unit[90]{\%} confidence level for $N_i$ decaying into generic 2-track final states for the six scenarios summarised in Table~\ref{tab:our benchmark scenarios} are shown in Figure~\ref{fig:result all scenarios}.
The region above the curves marks the expected exclusion regions for 2.3 events for each scenario.
The curves include the effect of the geometrical acceptance, and assume the background to be fully negligible.
Figures~\ref{fig:result e}--\subref{fig:result tau} show the NA62 sensitivity for the three extreme scenarios~D)--F) with a single dominant flavour compared to existing limits and projections of proposed experiments.

The sensitivity to particles of the hidden sector can be greatly reduced by the presence of background.
The interaction of the proton beam with the dump, along with the signals, give rise to copious short-lived resonances, as well as pions and kaons.
While the length of the dump ($\sim 22$ interaction lengths) is sufficient to absorb the hadrons and the electromagnetic radiation, the decays of pions, kaons and short-lived resonances result in a large flux of muons and neutrinos, which are the major sources of background in the apparatus.
In November 2016 during a $\sim 10$ hours long run in the dump mode using \unit[40]{\%} of the nominal beam intensity about $2\times 10^{15}$ POT were recorded and \unit[18]{kHz} of muons have been measured
within the NA62 acceptance.
A preliminary study of the background rates and topologies has been performed.
The analysis of this dataset shows that, within the current available statistics, NA62 can reduce the background to zero for all hidden sector particles decaying into fully reconstructed final states with two charged tracks.
In order to control the backgrounds of partially reconstructed final states or final states containing photons an upgrade of the current apparatus is required.
A detailed discussion of the background observed in NA62 when operated in the beam dump mode can be found in~\cite{Lanfranchi:2017wzl}.
Up to date, the dataset collected by NA62 in the beam-dump mode is about a factor 10 larger and the analysis is under progress.

\section{Discussion and Conclusion}

We have studied the sensitivity of the NA62 experiment in the dump mode to heavy neutral leptons in the low scale seesaw model.
The NA62 sensitivity was obtained for $10^{18}$ POT which is the data sample expected to be collected in 2021--2023.
This dataset corresponds to 3 months of data taking at full intensity and will be spread over said period of three years in order to allow NA62 to complete the kaon programme.
If the kaon programme will be completed earlier, more time can be dedicated to the beam dump mode, depending on the collaboration strategy.
At full intensity NA62 can collect up to $3\times 10^{18}$ POT per year.

Our main results are shown in Figure~\ref{fig:result all scenarios}.
Below the kaon mass, the accessible parameter space is already strongly constrained by past kaon and pion decay experiments.
Heavy neutrinos that are heavier than $D$-mesons can only be produced in $B$-meson decays at NA62, and their number is too small to give sizeable event rates.
As shown in Figures~\ref{fig:result e}--\subref{fig:result tau} NA62 is the only existing experiment that can probe significant fractions of unexplored parameter space for heavy neutrinos that are lighter than $D$-mesons.
Moreover, based on current estimates~\cite{Drewes:2016jae}, it appears to be the only existing experiment that can enter the parameter region where leptogenesis is possible in the $\nu$MSM.

\begin{figure}
\centering
\includegraphics[width=.618\linewidth]{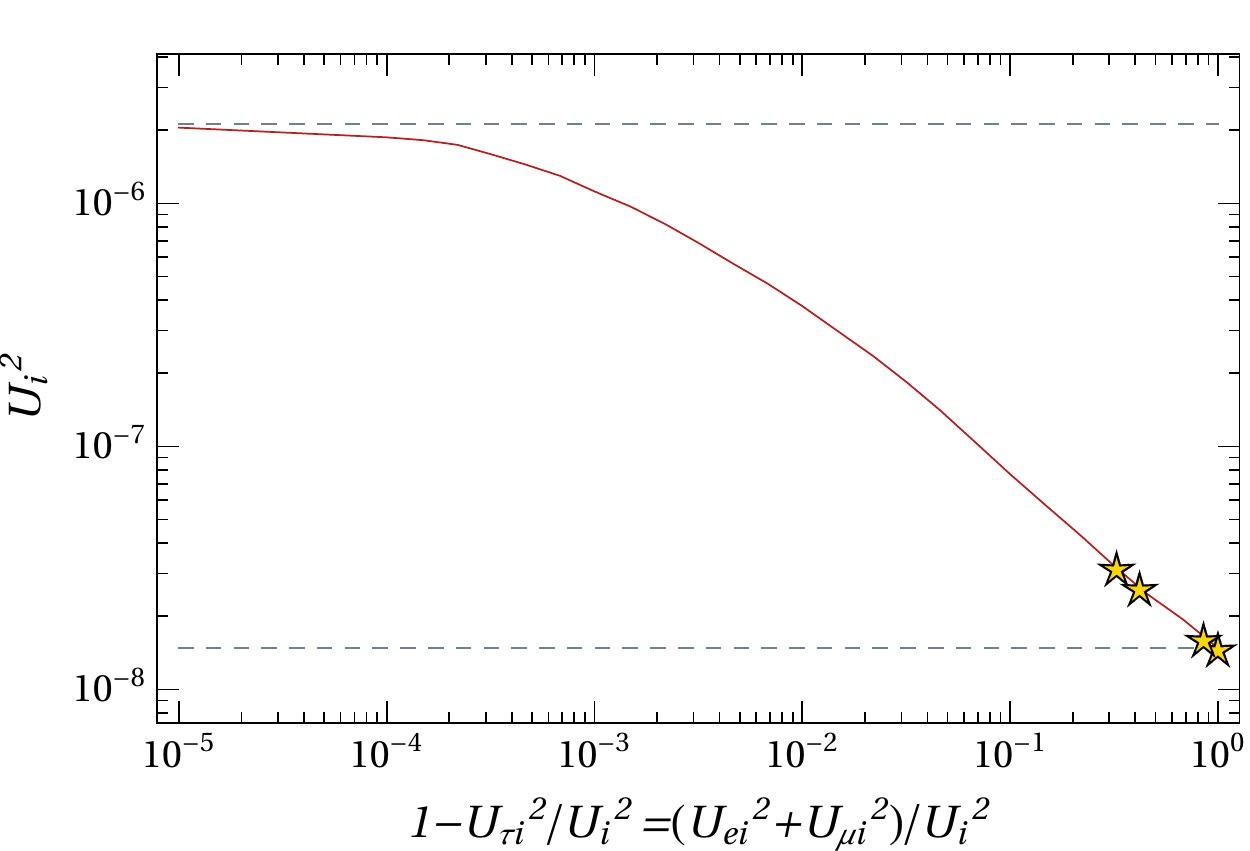}
\caption{Dependence of the NA62 sensitivity on $U_{\tau i}^2/U_i^2$ for $M_i = \unit[1.5]{GeV}$.
The stars indicate the benchmark scenarios A)--D), the horizontal lines mark the extreme scenarios with $U_i^2 = U_{e i}^2$ and $U_i^2 = U_{\tau i}^2$.
}
\label{fig:UtauDependence}
\end{figure}

As one can see in Figures~\ref{fig:result all scenarios} and~\ref{fig:UtauDependence}, the sensitivity can vary by about two orders of magnitude, depending on the "flavour mixing pattern", i.e., the way how the total mixing $U_i^2=\sum_a U_{ai}^2$ of a heavy neutrino $N_i$ is distributed amongst the different flavours $a = e,\:\mu,\:\tau$.
The primary reason is the tauon mass, which kinematically blocks the decay of $D$-mesons into $tau$ and
heavy neutrinos with masses above the kaon mass.
Hence, production of $N_i$ via $U_{\tau i}^2$ is only possible via $\tau$ decays and the decay of $N_i$ via $U_{\tau i}^2$ is only possible through neutral current interactions.
As a result, the sensitivity primarily depends on how $U_i^2$ is distributed amongst $U_{\tau i}^2$ and $U_{e i}^2+U_{\mu i}^2$.
For $M_i$ above the di-muon threshold, the sensitivity is essentially independent of $U_{e i}^2/U_{\mu i}^2$. This covers the entire mass range above the kaon mass, where NA62 can probe unexplored parameter space.
Remarkably, the sensitivity of NA62 in this regime shows a strong dependence on the flavour mixing pattern only if $U_{i}^2$ is dominated by $U_{\tau i}^2$, cf.~Figure~\ref{fig:UtauDependence}.

The flavour mixing pattern in the type-I seesaw model is strongly constrained by neutrino oscillation data.
These constraints depend on the number $n$ of heavy right handed neutrinos that generate the light neutrino masses.
For $n=2$ and $U_i^2$ that can be probed with NA62, global fits to current neutrino oscillation data constrain the ratios $U_{ai}^2/U_i^2$ to the filled areas in Figure~\ref{fig:AllowedAreas}.
These clearly exclude a strongly $U_{\tau i}^2$ dominated scenario.
Hence, the sensitivity in the minimal model with $n = 2$, which effectively also describes the $\nu$MSM, is rather independent of the flavour mixing pattern.
This statement can be quantified by comparing the scenarios A)--D) in Table~\ref{tab:our benchmark scenarios}, which mark the most extreme patterns that are allowed for $n = 2$ (cf.~stars in Figure~\ref{fig:AllowedAreas}).
They are represented by the first four lines in Figure~\ref{fig:result all scenarios}.

In Figures~\ref{fig:Chi2NO}--\ref{fig:Chi2IO} we perform a more detailed analysis to identify the likelihood for different flavour mixing patterns $U_{ai}^2/U_i^2$, based on present neutrino oscillation data.
Remarkably, these likelihoods only weakly depend on theoretical assumptions, i.e., directly reflect the statistical preference due to experimental data.
That is, the though the masses $M_i$ and overall coupling strengths $U_i^2$ of the hypothetical heavy neutrinos $N_i$ are unknown, current neutrino oscillation data already allows to make robust statements on the relative size of their couplings $U_{ai}^2$ to the different SM flavours.
This is the second main result of the present work, in addition to the NA62 sensitivity estimates in Figure~\ref{fig:result}.
In Figure~\ref{fig:Chi2NODUNE} we estimate how the constraints can improve if the DUNE experiment measures the Dirac phase $\delta$ in the light neutrino mixing matrix.
For $n=3$, mixing patterns outside the filled areas in Figure~\ref{fig:AllowedAreas} can be made consistent with neutrino oscillation data, including extreme $U_{\tau i}^2$ dominated scenarios.
However, these require considerable tunings in the parameters.
The preferred range of $U_{ai}^2/U_i^2$ in absence of tunings still roughly coincides with the filled regions in Figure~\ref{fig:AllowedAreas}.

In summary, we find that NA62 is the world's most powerful existing experiment to search for heavy neutrinos with masses between those of kaons and $D$-mesons.
In this mass range, the sensitivity within the parameter region that is preferred by neutrino oscillation data depends only weakly on the heavy neutrino flavour mixing pattern.

\subsection*{Acknowledgement}

We would like to thank Eduardo Cortina Gil, Bj\"orn Garbrecht, Dario Gueter and Babette D\"obrich for very helpful discussions.
This research was supported by the DFG cluster of excellence "Origin and Structure of the Universe" (\url{universe-cluster.de}) and by the Collaborative Research Center SFB1258 of the Deutsche Forschungsgemeinschaft.
Jan Hajer is supported by the Research Grants Council of the Hong Kong S.A.R.\ under the Collaborative Research Fund (CRF) Grant \textnumero\ HUKST4/CRF/13G.

\appendix

\section{Heavy neutrino mixing angles}
\label{App:Mixing}

Here we present the relation between $U_{ai}^2$ and the parameters in $U_\nu$ for $n = 2$ as found in~\cite{Drewes:2016jae}.
We use the shorthand notations $s_{ab} = \sin\uptheta_{ab}$ and $c_{ab} = \cos\uptheta_{ab}$.%
\footnote{More precisely, we take $s_{ab}$ and $c_{ab}$ to be the positive real roots of $s_{ab}^2$ and $c_{ab}^2$ from Table~\ref{tab:active_bounds}.}
It is worth mentioning the limit
\begin{align}
\label{eq:limits}
\lim_{\upepsilon\to0}\tanh(2\Im\omega)& = 1\ ,&\lim_{\upepsilon\to0}\cosh(2\Im\omega)& = \sinh(2\Im\omega) = \frac12 \exp(2\Im\omega) = \frac{1}{2\upepsilon}\ ,
\end{align}
which is helpful to obtain the symmetry protected limit from the following equations.

\paragraph{Normal hierarchy}

\begin{subequations}
\begin{align}
    2 M_{1,2} U^2_{e1,2}
=&\ a_1^+ \cosh(2\Im\omega) - a_2 \sin \left(\frac{\alpha_2}{2} + \delta\right) \sinh(2 \Im \omega) \notag \\
&\pm \left[a_1^- \cos(2 \Re \omega) - a_2 \sin \left(\frac{\alpha_2}{2} + \delta\right) \sin(2 \Re \omega)\right]
\ ,\\
    2 M_{1,2} U^2_{\mu 1,2}
=&\ \left[a_3^+ -a_4\cos(\delta)\right] \cosh(2 \Im \omega)-\left[a_5\sin\left(\frac{\alpha_2}{2}\right)-a_6\sin\left(\frac{\alpha_2}{2}+\delta\right)\right]\sinh(2\Im\omega)\notag\\
&\mp \left[ a_3^-+a_4 \cos(\delta) \right] \cos(2 \Re \omega) \mp \left[a_5\cos\left(\frac{\alpha_2}{2}\right)-a_6\cos\left(\frac{\alpha_2}{2}+\delta\right)\right]\sin(2\Re\omega)
\ ,\\
2 M_{1,2} U^2_{\tau 1,2} = &\ \left[a_7^+ +a_4\cos(\delta)\right] \cosh(2 \Im \omega) + \left[a_5\sin\left(\frac{\alpha_2}{2}\right)+a_{8}\sin\left(\frac{\alpha_2}{2}+\delta\right)\right]\sinh(2\Im\omega)\notag\\
&\mp \left[a_7^--a_4\cos(\delta)\right] \cos(2\Re\omega) \pm \left[a_5\cos\left(\frac{\alpha_2}{2}\right)+a_{8}\cos\left(\frac{\alpha_2}{2}+\delta\right)\right]\sin(2\Re\omega)
\ ,
\end{align}
\label{eq:mixing_NOv2}
\end{subequations}
with $a_1$ to $a_8$ positive real values that are given by active neutrino masses and their mixing angles
\begin{subequations}
\begin{align}
a_1^\pm & = m_2c_{13}^2s_{12}^2\pm m_3s_{13}^2\ ,\\
a_2& = 2\sqrt{m_2m_3}c_{13}s_{12}s_{13}\xi\ ,\\
a_3^\pm & = \pm m_2(c_{12}^2c_{23}^2+s_{12}^2s_{13}^2s_{23}^2)
 + m_3c_{13}^2s_{23}^2\ ,\\
a_4& = 2m_2c_{12}c_{23}s_{12}s_{13}s_{23}\ ,\\
a_5& = 2\sqrt{m_2m_3}c_{12}c_{13}c_{23}s_{23}\xi\ ,\\
a_6& = 2\sqrt{m_2m_3}c_{13}s_{12}s_{13}s_{23}^2\xi\ ,\\
a_7^\pm & = \pm m_2(c_{23}^2s_{12}^2s_{13}^2+c_{12}^2s_{23}^2)
+ m_3c_{13}^2c_{23}^2\ ,\\
a_{8}& = 2\sqrt{m_2m_3}c_{13}c_{23}^2s_{12}s_{13}\xi\ .
\end{align}
\end{subequations}
Here we have set $\alpha_1 = 0$ without loss of generality, which allows to use the simplified notation $\alpha_2 = \alpha$ that we adopt in the main text.

\paragraph{Inverted hierarchy}

\begin{subequations}
\begin{align}
2 M_{1,2} U^2_{e1,2} = &\ b_1^+ \cosh(2\Im\omega) +b_2\sin\left(\frac{\widetilde \alpha}{2}\right)\sinh(2\Im\omega)\notag\\
&\pm b_1^- \cos(2\Re\omega) \mp b_2\cos\left(\frac{\widetilde \alpha}{2}\right)\sin(2\Re\omega)
\ ,\\
2 M_{1,2} U^2_{\mu1,2} = &\
\left[b_3^+ -b_4^+\cos(\delta)\right] \cosh(2\Im\omega)
\mp \left[b_3^- +b_4^-\cos(\delta)\right] \cos(2\Re\omega) \notag\\
&-\left[b_5\sin\left(\frac{\widetilde \alpha}{2}\right)+b_6\sin\left(\frac{\widetilde \alpha}{2} - \delta\right)-b_7\sin\left(\frac{\widetilde \alpha}{2} + \delta\right)\right]\sinh(2\Im\omega) \notag\\
&\pm \left[b_5\cos\left(\frac{\widetilde \alpha}{2}\right)+b_6\cos\left(\frac{\widetilde \alpha}{2} - \delta\right) - b_7\cos\left(\frac{\widetilde \alpha}{2} + \delta\right)\right]\sin(2\Re\omega)
\ ,\\
2 M_{1,2} U^2_{\tau 1,2} = &\
\left[ b_8^+ + b_4^+ \cos(\delta)\right] \cosh(2\Im\omega)
\mp \left[b_8^- - b_4^ - \cos(\delta)\right]\cos(2\Re\omega) \notag \\
&-\left[b_9\sin\left(\frac{\widetilde \alpha}{2}\right) - b_{6}\sin\left(\frac{\widetilde \alpha}{2}-\delta\right) + b_7\sin\left(\frac{\widetilde \alpha}{2}+\delta\right)\right]\sinh(2\Im\omega) \notag \\
&\pm\left[b_9\cos\left(\frac{\widetilde \alpha}{2}\right) - b_{6}\cos\left(\frac{\widetilde \alpha}{2}-\delta\right) + b_7\cos\left(\frac{\widetilde \alpha}{2} + \delta\right)\right]\sin(2\Re\omega)
\ ,
\end{align}
\label{eq:mixing_IOv2}
\end{subequations}
with $b_1$ to $b_{9}$ positive real numbers given by the active neutrino masses and mixings
\begin{subequations}
\begin{align}
b_1^\pm& = m_1c_{12}^2c_{13}^2\pm m_2s_{12}^2c_{13}^2\ ,\\
b_2& = 2\sqrt{m_1m_2}c_{12}s_{12}\xi\ ,\\
b_3^\pm& = \pm m_1(c_{23}^2s_{12}^2+c_{12}^2s_{13}^2s_{23}^2)
 + m_2(c_{12}^2c_{23}^2+s_{12}^2s_{13}^2s_{23}^2)\ ,\\
b_4^\pm & = 2(\pm m_2 -m_1)c_{12}c_{23}s_{12}s_{13}s_{23}\ ,\\
b_5& = 2\sqrt{m_1m_2}(c_{12}c_{23}^2s_{12}-c_{12}s_{12}s_{13}^2s_{23}^2)\xi\ ,\\
b_6& = 2\sqrt{m_1m_2}c_{12}^2c_{23}s_{13}s_{23}\xi\ ,\\
b_7& = 2\sqrt{m_1m_2}s_{12}^2c_{23}s_{13}s_{23}\xi\ ,\\
b_8^\pm& = \pm m_1(c_{12}^2 c_{23}^2 s_{13}^2 +s_{12}^2 s_{23}^2)
+ m_2(c_{23}^2s_{12}^2 s_{13}^2 + c_{12}^2 s_{23}^2 )\ ,\\
b_9& = 2\sqrt{m_1m_2}c_{12}s_{12}(s_{23}^2-c_{23}^2s_{13}^2)\xi\ ,
\end{align}
\end{subequations}
and $\widetilde \alpha = \alpha_2-\alpha_1$.
This implies that for inverted hierarchy the Yukawa matrices $F$ only depend on the difference $\alpha_2-\alpha_1$, which allows us to set $\alpha_1 = 0$ and use the simplified notation $\alpha_2 = \widetilde \alpha = \alpha$.

\printbibliography

\end{document}